\documentclass[aps,prl,twocolumn,superscriptaddress]{revtex4-1}
\usepackage{amsmath,amsthm,amssymb}
\usepackage[dvipdfmx]{graphicx}
\usepackage{amsmath,bm}
\usepackage{color,ulem}
\usepackage{ifthen}

\newcount \commentout
\newcommand{\1}{\mbox{1}\hspace{-0.25em}\mbox{l}}

\newlength{\figwidth}
\newlength{\figlarge}
\begin{document}
%%%%%%%%%%%%%%%%%%%%%%%%%%%%%%%%%%%%%%%%%%%%%%%%%%%%%%%%%%%%%%%%%%%%%%%
\title{
Exceptional rings in two-dimensional correlated systems with chiral symmetry
}
%%%%%%%%%%%%%%%%%%%%%%%%%%%%%%%%%%%%%%%%%%%%%%%%%%%%%%%%%%%%%%%%%%%%%%%
\author{Tsuneya Yoshida}
\affiliation{Department of Physics, University of Tsukuba, Ibaraki 305-8571, Japan}
\author{Robert Peters}
\affiliation{Department of Physics, Kyoto University, Kyoto 606-8502, Japan}
\author{Norio Kawakami}
\affiliation{Department of Physics, Kyoto University, Kyoto 606-8502, Japan}
\author{Yasuhiro Hatsugai}
\affiliation{Department of Physics, University of Tsukuba, Ibaraki 305-8571, Japan}
%%%%%%%%%%%%%%%%%%%%
%%%%%%%%%%%%%%%%%%%%%%%%%%%%%%%%%%%%%%%%%%%%%%%%%%%%%%%%%%%%%%%%%%%%%%%
\date{\today}
%%%%%%%%%%%%%%%%%%%%%%%%%%%%%%%%%%%%%%%%%%%%%%%%%%%%%%%%%%%%%%%%%%%%%%%
\begin{abstract}
Emergence of exceptional points in two dimensions is one of the remarkable phenomena in non-Hermitian systems. 
We here elucidate the impacts of symmetry on the non-Hermitian physics.
Specifically, we analyze chiral symmetric correlated systems in equilibrium where the non-Hermitian phenomena are induced by the finite lifetime of quasi-particles.
Intriguingly, our analysis reveals that the combination of symmetry and non-Hermiticity results in novel topological degeneracies of energy bands which we call symmetry-protected exceptional rings (SPERs). 
We observe the emergence of SPERs by analyzing a non-Hermitian Dirac Hamiltonian. Furthermore, by employing the dynamical mean-field theory, we demonstrate the emergence of SPERs in a correlated honeycomb lattice model whose single-particle spectrum is described by a non-Hermitian Dirac Hamiltonian. We uncover that the SPERs survive even beyond the non-Hermitian Dirac Hamiltonian, which is related to a zero-th Chern number.
\end{abstract}
%%%%%%%%%%%%%%%%%%%%%%%%%%%%%%%%%%%%%%%%%%%%%%%%%%%%%%%%%%%%%%%%%%%%%%%
\pacs{
***
} 
%%%%%%%%%%%%%%%%%%%%%%%%%%%%%%%%%%%%%%%%%%%%%%%%%%%%%%%%%%%%%%%%%%%%%%%
%%%%%%%%%%%%%%%%%%%%%%%%%%%%%%%%%%%%%%%%%%%%%%%%%%%%%%%%%%%%%%%%%%%%%%%
\maketitle
%%%%%%%%%%%%%%%%%%%%%%%%%%%%%%%%%%%%%%%%%%%%%%%%%%%%%%%%%%%%%%%%%%%%%%%

%%%%%%%%%%%%%%%%%%
\textit{Introduction.--}
%%%%%%%%%%%%%%%%%%
In these decades, the topological perspective on condensed matter systems increases its importance~\cite{TI_review_Hasan10,TI_review_Qi10,Ando_review_JPSJ13}. 
One of the significant advances in this field is that the notion of topology for gapped systems is extended to gapless systems. 
A representative example is the Weyl fermion emerging in the bulk of TaAs~\cite{Weng_TaAs_LDA_PRX15,Xu_TaAs_arpes_Sci15} which is topologically protected by a nontrivial change of the Chern numbers~\cite{Weyl_theor_PRB11,Burkov_Weyl_theor_PRL11}.
These low energy states induce Fermi arcs at the surfaces of the materials, which results in a finite value of the Hall conductivity.
Another important advance in this field is the discovery of topological properties protected by symmetry.
Symmetry imposes constraints on wave functions, resulting in enriched topological properties protected by the symmetry. 
For instance, time-reversal symmetry protects the topology of the celebrated $\mathbb{Z}_2$ topological insulators~\cite{Kane_PRL05_1,Kane_PRL05_2,Bernevig_QSH_Science06} realized in a quantum well of $\mathrm{HgTe}/\mathrm{CdTe}$~\cite{Konig_Science07}. 
Particle-hole symmetry protects the topology of superconductors whose gapless edge states play a crucial role in the emergence of Majorana fermions~\cite{Kitaev_chain_01,Majorana_Ryu_RPL02,Alicia_Majorana_review12,Majorana_Mourik,Majorana_Rokhinson2012,Majorana_Das2012,Sato_JPSJ16}.
Such diversification of topological systems is quite ubiquitous. 
Symmetry also enriches the topological gapless excitations; the combination of time-reversal symmetry and spatial symmetry protects $\mathbb{Z}_2$ topological semi-metals~\cite{Morimoto_Weyl_RPB14,Zhao_PRL16}.
Thanks to the combination of topology and symmetry, a variety of topological degeneracies of energy bands have been discovered so far.

Along with the above great success of topological phases, a new arena of condensed matter physics has been pioneered; systems described by non-Hermitian Hamiltonians have been realized in the presence of dissipation or under continuous observation,~\textit{etc.}~\cite{Hatano-NelsonPRL96,Fukui-KawakamiPRB98,CMBender_PRL98,Guo_nHExp_PRL09,Ruter_nHExp_NatPhys10,Regensburger_nHExp_Nat12,Ashida_PRA16,YAshida_NatCom17}. 
Furthermore, it turned out that even ordinary strongly correlated systems in equilibrium provide a platform of the non-Hermitian physics because of the lifetime of quasi-particles~\cite{VKozii2017_non-Hermi}. 
Namely, in general, the structure of the single-particle spectrum for the frequency $\omega$ and the momentum $\bm{k}$ is governed by the non-Hermitian effective Hamiltonian defined as $H_{eff}(\omega,\bm{k}) :=h(\bm{k}) +\Sigma(\omega+i\delta,\bm{k})$, where the Hermitian matrix $h(\bm{k})$ describes the one-body part of the Hamiltonian, and $\Sigma(\omega+i\delta,\bm{k})$ denotes the self-energy with an infinitesimal positive constant $\delta$.
One of the remarkable properties for such non-Hermitian systems is the emergence of the novel topological degeneracy arising from non-Hermiticity of the effective Hamiltonian. 
At certain points of the Brillouin zone (BZ) the effective Hamiltonian becomes defective, inducing the degeneracy of the energy bands~\cite{Zhen_EP_Nature15,HShen2017_non-Hermi,YXuPRL17_exceptional_ring}. 
Interestingly, these points are accompanied by robust Fermi arcs in the bulk where the energy gap becomes pure imaginary~\cite{VKozii2017_non-Hermi,Yoshida_nHKondo_PRB18}.

The above discovery of the new topological degeneracy implies that non-Hermitian systems potentially host a variety of novel topological phenomena which cannot be observed for Hermitian systems~\cite{CMBender_PRL98,EsakiSatoHasebeKhomoto_12,SatoEsakiHasebeKhomoto_12,San-Jose2016,TELeePRL16_Half_quantized,ZPGong_PRL17,Zyuzin_disorder_WeylPRB18,ZPGong_PRX18,KKawabata_arXiv18,Yao_1DBBC_PRL2018,Yao_2DBBC_PRL2018}. 

In particular, with the above mentioned great success for Hermitian systems, one can easily expect that non-Hermitian systems host a diversity of topological degeneracies in the presence of symmetry. The discovery of symmetry-protected topological degeneracies would pioneer a new direction of research in topological systems.
Despite this significant issue, however, few studies have addressed exceptional points by taking into account symmetry, so far.

We here study impacts of symmetry on the topological degeneracies arising from non-Hermiticity. In particular, we analyze correlated systems with chiral symmetry from the non-Hermitian perspective.
Remarkably, our analysis discovers novel topological degeneracies in two-dimensional systems which we call symmetry-protected exceptional rings (SPERs). 
Specifically, we uncover the emergence of SPERs based on the following results. 
For non-Hermitian Dirac models, we show that chiral symmetry for correlated systems protects the exceptional rings. 
By employing the dynamical mean-field theory~\cite{WMetznerPRL89_DMFT,MHartmannZP89_DMFT,AGeorgesRMP96_DMFT} with the numerical renormalization group~\cite{KWilsonRMP75_NRG,RPetersPRB06_NRG,RBullaRMP08_NRG} (DMFT+NRG), we also demonstrate that a honeycomb lattice model with inhomogeneous Hubbard interactions hosts the SPERs. The single-particle spectrum of this model is essentially reduced to the non-Hermitian Dirac Hamiltonian with $2\times 2$ matrices.
Furthermore, we show that the exceptional rings survive even beyond the Dirac models. This remarkable behavior is related to a zero-th Chern number.

%%%%%%%%%%%%%%%%%%
\textit{
Extended chiral symmetry arising from symmetry of the many-body Hamiltonian.--
}
%%%%%%%%%%%%%%%%%%
We first discuss a constraint on the non-Hermitian effective Hamiltonian imposed by the chiral symmetry for many-body systems.
Consider a correlated system with chiral symmetry. Then, the many-body Hamiltonian 
$\hat{H}$ satisfies $\hat{U}^\dagger_\Gamma \hat{H}^* \hat{U}_\Gamma =\hat{H}$~\cite{Hatsugai_mb_chiral_JPSJ06,Gurarie_PRB11}. 
Here, $\hat{\Gamma}:=\hat{U}_\Gamma\mathcal{K}$ denotes the second quantized chiral operator with $\hat{U}^2_\Gamma=1$. 
The operator $\mathcal{K}$ takes complex conjugation. The chiral operator transforms
the annihilation operator $c_{jn}$ as follows: 
$\hat{U}^\dagger_\Gamma \hat{c}_{jn} \hat{U}_\Gamma = \hat{c}^\dagger_{jm}U^\dagger_{\Gamma,mn}$, 
where $c_{jn}$ annihilates a fermion at site $j$, and $n$ labels the internal degrees of freedom such as sublattice, spin, orbital, \textit{etc.}
The matrix $U_{\Gamma}$ is the chiral operator of the non-interacting Hamiltonian $h_{ij}$ satisfying $U^\dagger_{\Gamma} h_{ij} U_{\Gamma}=-h_{ij}$ with $U^2_{\Gamma}=\1$. Here $\1$ denotes the identity matrix. 
The explicit form of the second quantized chiral operator $\hat{\Gamma}$ is defined in Eq.~(\ref{eq: hatUgamma_honey}) for systems composed of two-sublattices.

As discussed in Sec.~\ref{sec: chi_symm_GF} of the supplemental material~\cite{suppl}, the many-body chiral symmetry results in the following constraint on the non-Hermitian effective Hamiltonian: $H_{eff}(\omega,\bm{k})=- U^\dagger_\Gamma H^\dagger_{eff}(-\omega,\bm{k}) U_\Gamma$.
In particular, at $\omega=0$, this constraint is reduced to
%%%%%%%%%%%%%%%%%%
\begin{eqnarray}
\label{eq: chiral_Heff_main}
 H_{eff}(0,\bm{k})  &=& - U^\dagger_\Gamma H^\dagger_{eff}(0,\bm{k}) U_\Gamma,
\end{eqnarray}
%%%%%%%%%%%%%%%%%%
which we call extended chiral symmetry.
In the following, we show that the extended chiral symmetry protects the exceptional rings appearing in the single-particle spectrum. As a first step, we clarify the emergence of SPERs for non-Hermitian Dirac Hamiltonians.
%

%%%%%%%%%%%%%%%%%%
\textit{
Generic argument of non-Hermitian Dirac models.--
}
%%%%%%%%%%%%%%%%%%
Here, we discuss a generic non-Hermitian Dirac Hamiltonian with extended chiral symmetry in two dimensions. 
We suppose that the effective Hamiltonian $H_{eff}(0,\bm{k})$ can be expanded by basis elements of Clifford algebra $\gamma_\alpha$ ($\alpha=1,\cdots,2l-1$ with $l\in \mathbb{Z}$), satisfying $\gamma^\dagger_\alpha=\gamma_\alpha, \ \gamma^2_\alpha=\1$, and $\{ \gamma_\alpha,\gamma_\alpha'\}=0$ for $\alpha\neq \alpha'$. 
For $l=2$ ($l=3$), $\gamma$'s correspond to the Pauli (Dirac) matrices, respectively. As discussed later, the SPERs are robust even beyond Dirac models. 

Let us start with the case where no symmetry is imposed on the non-Hermitian effective Hamiltonian.
In this case, the effective Hamiltonian is expanded as
%%%%%%%%%%%%%%%%%
\begin{eqnarray}
\label{eq: no symm dirac}
 H_{eff}(0,\bm{k}) &=& b_\alpha(\bm{k}) \gamma_\alpha + i d_{\alpha}\gamma_{\alpha}(\bm{k}),
\end{eqnarray}
%%%%%%%%%%%%%%%%%%
with $b_\alpha$ and $d_{\alpha}$ ($\alpha=0,\cdots,2l-1$) taking real numbers. Here summation is assumed over repeated indices $\alpha=0,1,\cdots,2l-1$. $\gamma_0$ is the identity operator.
We note that $\left[b_{\mu}(\bm{k}) \gamma_{\mu} + i d_{\mu}(\bm{k})\gamma_{\mu}  \right]^2=b^2-d^2+i2\bm{b}\cdot\bm{d}$ holds with $b^2:=b_{\mu} b_{\mu}$, $d^2:=d_{\mu} d_{\mu}$, and $\bm{b}\cdot\bm{d}:= b_{\mu} d_{\mu}$, respectively. 
Here, we have assumed summation over $\mu=1,\cdots 2l-1$. 
Thus, eigenvalues of the above Hamiltonian are written as
%%%%%%%%%%%%%%%%%%
\begin{eqnarray}
E_{\pm}(\bm{k}) &=& b_0(\bm{k})+id_0(\bm{k})\pm \sqrt{b^2(\bm{k})-d^2(\bm{k})+2i\bm{b}(\bm{k})\cdot\bm{d}(\bm{k})}, \nonumber 
\end{eqnarray}
%%%%%%%%%%%%%%%%%%
%
An exceptional point emerges at $\bm{k}_0$ in the BZ satisfying the following two conditions:
%%%%%%%%%%%%%%%%%%
\begin{eqnarray}
\label{eq: cond_EP_nosymm_dirac}
 b^2(\bm{k})-d^2(\bm{k})=0, &\quad& \bm{b}(\bm{k})\cdot\bm{d}(\bm{k})=0,
\end{eqnarray}
%%%%%%%%%%%%%%%%%%
which can be seen as follows. In a proper basis, the effective Hamiltonian can be rewritten as 
%%%%%%%%%%%%%%%%%%
\begin{eqnarray}
 H_{eff}(0,\bm{k}_0) &=& b[\sigma_1\otimes A + i\sigma_2\otimes\1],
\end{eqnarray}
%%%%%%%%%%%%%%%%%%
where $\1$ denotes the identity matrix, $A$ is a Hermitian matrix, and $\sigma$'s denote the Pauli matrices. 
One can easily see that the effective Hamiltonian $H_{eff}(0,\bm{k}_0)$ cannot be diagonalized with any unitary matrix.

Now, let us show how the extended chiral symmetry protects the exceptional rings in two dimensions. 
By choosing $\gamma_{2l-1}$ as the chiral matrix, we can see that the following parameters vanish due to extended chiral symmetry~(\ref{eq: chiral_Heff_main}):
%%%%%%%%%%%%%%%%%%
\begin{eqnarray}
 b_0 = b_{2l-1} =d_{1}=\cdots=d_{2l-2}=0.
\end{eqnarray}
%%%%%%%%%%%%%%%%%%
Thus, the second condition of Eq.~(\ref{eq: cond_EP_nosymm_dirac}) is satisfied automatically by the symmetry, which is a key ingredient for SPERs.
The first condition is satisfied in the region where the energy band $b(\bm{k})$ for the Hermitian Dirac Hamiltonian intersects another manifold described by the continuous function $\left| d_{2l-1}(\bm{k})\right|$.
Noticing that each of them forms a two-dimensional manifold parameterized by the two-dimensional BZ, we can conclude that SPERs are allowed as closed one-dimensional manifolds for the non-Hermitian Dirac Hamiltonian in the presence of extended chiral symmetry.

We note that the SPERs in the two-dimensional BZ cannot be characterized by the vorticity introduced in Ref.~\onlinecite{HShen2017_non-Hermi}. This is because the computation of the vorticity requires a closed path surrounding the exceptional points, while there is no such a path in the two-dimensional BZ in the case of SPERs.
It is also worth to note that the above argument can be extended to three-dimensional systems. In such systems, the extended chiral symmetry results in exceptional spheres which are closed two-dimensional manifolds in the BZ.

In the above, we have seen the emergence of SPERs when the non-Hermitian effective Hamiltonian can be expanded in basis elements of Clifford algebra. 
In the following, we demonstrate the emergence of SPERs for a Hubbard model whose single-particle spectrum is described by the above non-Hermitian Dirac Hamiltonian with $l=2$.

\begin{figure}[!h]
\begin{minipage}{0.75\hsize}
\begin{center}
\includegraphics[width=\hsize,clip]{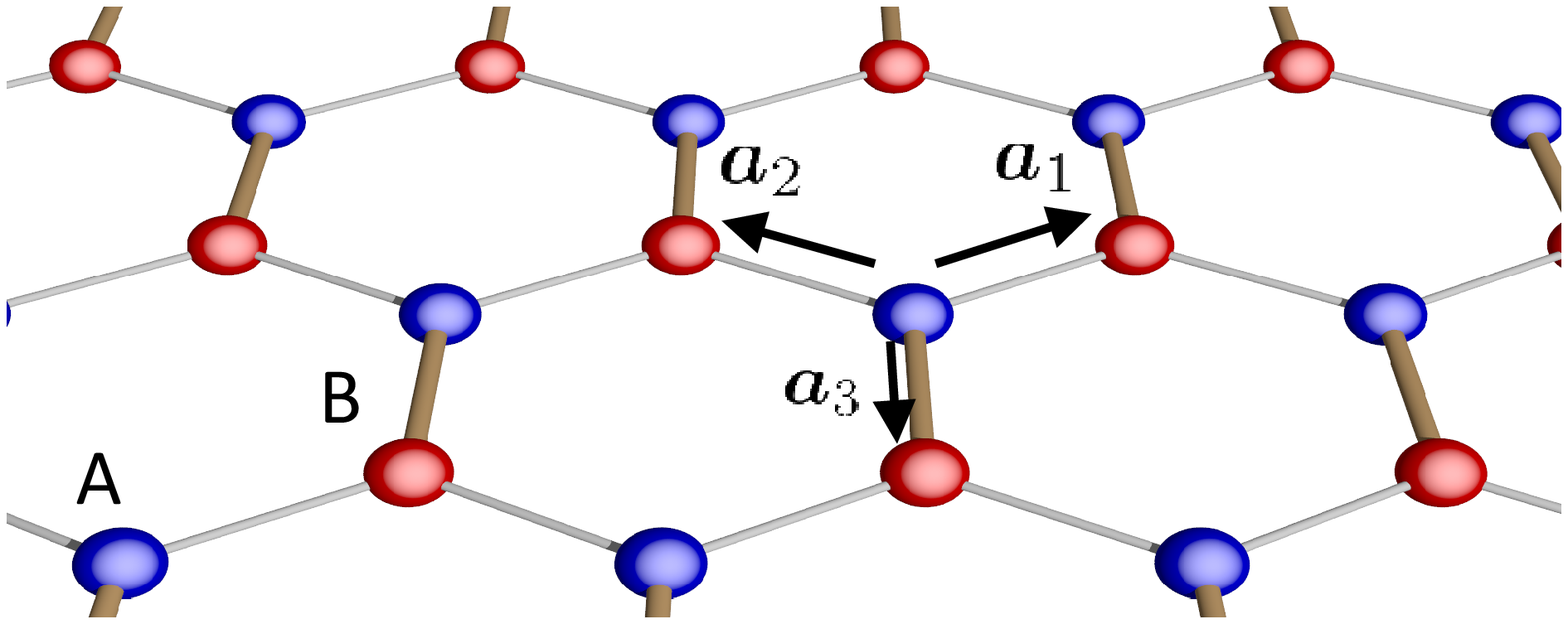}
\end{center}
\end{minipage}
\caption{(Color Online).
Sketch of the honeycomb lattice. Blue (red) spheres denote the $A$- ($B$-) sublattice, respectively. $\bm{a}_1$, $\bm{a}_2$, and $\bm{a}_3$ denote vectors connecting neighboring sites 
which are defined as 
$\bm{a}_1:=(\sqrt{3},1)/2$, $\bm{a}_2:=(-\sqrt{3},1)/2$, and $\bm{a}_3:=(0,-1)$. 
Gray (brown) bonds represent hopping $t$ ($rt$), respectively.
}
\label{fig: model}
\end{figure}

%%%%%%%%%%%%%%%%%%
\textit{
SPERs in a correlated honeycomb lattice.--
}
%%%%%%%%%%%%%%%%%%
Here, we demonstrate the emergence of SPERs in two dimensions by applying DMFT+NRG to the following honeycomb lattice model with inhomogeneous Hubbard interactions;
%%%%%%%%%%%%%%%%%%
\begin{eqnarray}
\label{eq: Hami}
\hat{H} &=& \sum_{\langle is, js'\rangle \sigma} t_{is,js'} \hat{c}^\dagger_{i s\sigma} \hat{c}_{j s'\sigma} +\sum_{i s}U_s (\hat{n}_{is\uparrow}-\frac{1}{2})(\hat{n}_{is\downarrow}-\frac{1}{2}), \nonumber 
\end{eqnarray}
%%%%%%%%%%%%%%%%%%
where $\hat{c}^\dagger_{i s\sigma}$ creates a fermion in the spin-state $\sigma=\uparrow,\downarrow$ at sublattice $s=A,B$ of site $i$. $\hat{n}_{is}=\hat{c}^\dagger_{i s\sigma}\hat{c}_{i s\sigma}$. 
The first term of the above Hamiltonian describes the hopping of the fermions between neighboring sites; its amplitude $t_{is,js'}$ takes $t \in \mathbb{R}$ for the direction of $\bm{a}_1$ or $\bm{a}_2$, while it takes $rt$ with $r \in \mathbb{R}$ for the direction of $\bm{a}_3$~(see Fig.~\ref{fig: model}).
The second term describes inhomogeneous Hubbard interactions with $U_s\in \mathbb{R}$.
%%%
This model is considered to be fabricated for cold atoms.
We note that the honeycomb lattice model has already been realized for cold atoms~\cite{Becker_honey_coldIOP10}. 
The inhomogeneous Hubbard interactions are considered to be introduced by employing the optical Feshbach resonance~\cite{RYamazaki_PRL10,LWClark_PRL15}.
%%5

Before presenting the DMFT results, we discuss the symmetry of this model and show that the effective Hamiltonian corresponds to the non-Hermitian Dirac Hamiltonian with $l=2$ discussed above.
For two-sublattice models, the second quantized chiral operator is given by
%%%%%%%%%%%%%%%%%%
\begin{eqnarray}
\label{eq: hatUgamma_honey}
 \hat{\Gamma} \!&=&\! \prod_{js} \left( \hat{c}^\dagger_{js\uparrow}\!+\mathrm{sgn}(s)\hat{c}_{js\uparrow}\right)\!\left( \hat{c}^\dagger_{js\downarrow}\!+\mathrm{sgn}(s)\hat{c}_{js\downarrow}\right)\!\mathcal{K},
\end{eqnarray}
%%%%%%%%%%%%%%%%%%
where $\mathrm{sgn}(s)$ takes $1$ and $-1$ for $s=A$ and $s=B$, respectively~\cite{Hatsugai_mb_chiral_JPSJ06,Gurarie_PRB11,Manmana_PRB12,TMI_YoshidaPRL14,Yoshida_ZtoZ4_PRL18}. 
We note $\hat{U}_\Gamma^2=1$ holds. 
Since our model preserves the $z$-component of the total spin, the effective Hamiltonian $H_{eff}(0,\bm{k})$ is expanded by the Pauli matrices $\tau$'s as follows:
$H_{eff}(0,\bm{k}) = id_0(\bm{k})\tau_0+\left[\bm{b}(\bm{k})+i\bm{d}(\bm{k})\right]\cdot \bm{\tau}$ 
with $\bm{b}(\bm{k}):=\left(b_1(\bm{k}), b_2(\bm{k}), 0 \right)$ and $\bm{d}(\bm{k}):=\left(0,0,d_3(\bm{k}) \right)$, which are given by~\cite{footnote_bDMFT}
%%%%%%%%%%%%%%%%%%
\begin{eqnarray}
\label{eq: bs_honeycomb}
b_1(\bm{k}) + ib_2(\bm{k})&=& t\sum_{j=1,2}e^{i\bm{k}\cdot\bm{a}_j}+rte^{i\bm{k}\cdot\bm{a}_3}, \\
d_0(\bm{k})\tau_0+d_3(\bm{k})\tau_3&=& \mathrm{Im} \Sigma(i\delta,\bm{k}).
\end{eqnarray}
%%%%%%%%%%%%%%%%%%
Here, the Pauli matrices $\tau$'s act on the sublattice space. 
We note that the chiral matrix is written as $U_\Gamma=\tau_3$. As shown above, the single-particle spectrum of this model is governed by the non-Hermitian Dirac Hamiltonian with $l=2$.

%%%%%%%%%%%%%%%%%%%%%%%%%
%%%%%%%%%%%%%%%%%%%%%%%%%

Now, we analyze the above model with inhomogeneous Hubbard interactions by applying the DMFT+NRG~\cite{WMetznerPRL89_DMFT,MHartmannZP89_DMFT,AGeorgesRMP96_DMFT,KWilsonRMP75_NRG,RPetersPRB06_NRG,RBullaRMP08_NRG} which treats local correlations exactly.
In order to treat inhomogeneity with the DMFT framework, we employ the sublattice method~\cite{AGeorgesRMP96_DMFT,Yoshida_DMFT_PRB12,Yoshida_PRB13_AFTI}.
%%%%%%%%%%%%%%%%%%
We first note that this model shows a first order Mott transition for $U_{Ac}\sim 11.5t$ at $T=0.05t$. Here, we set $U_B=U_A/2$.

\begin{figure}[!h]
\begin{minipage}{1\hsize}
\begin{center}
\includegraphics[width=\hsize,clip]{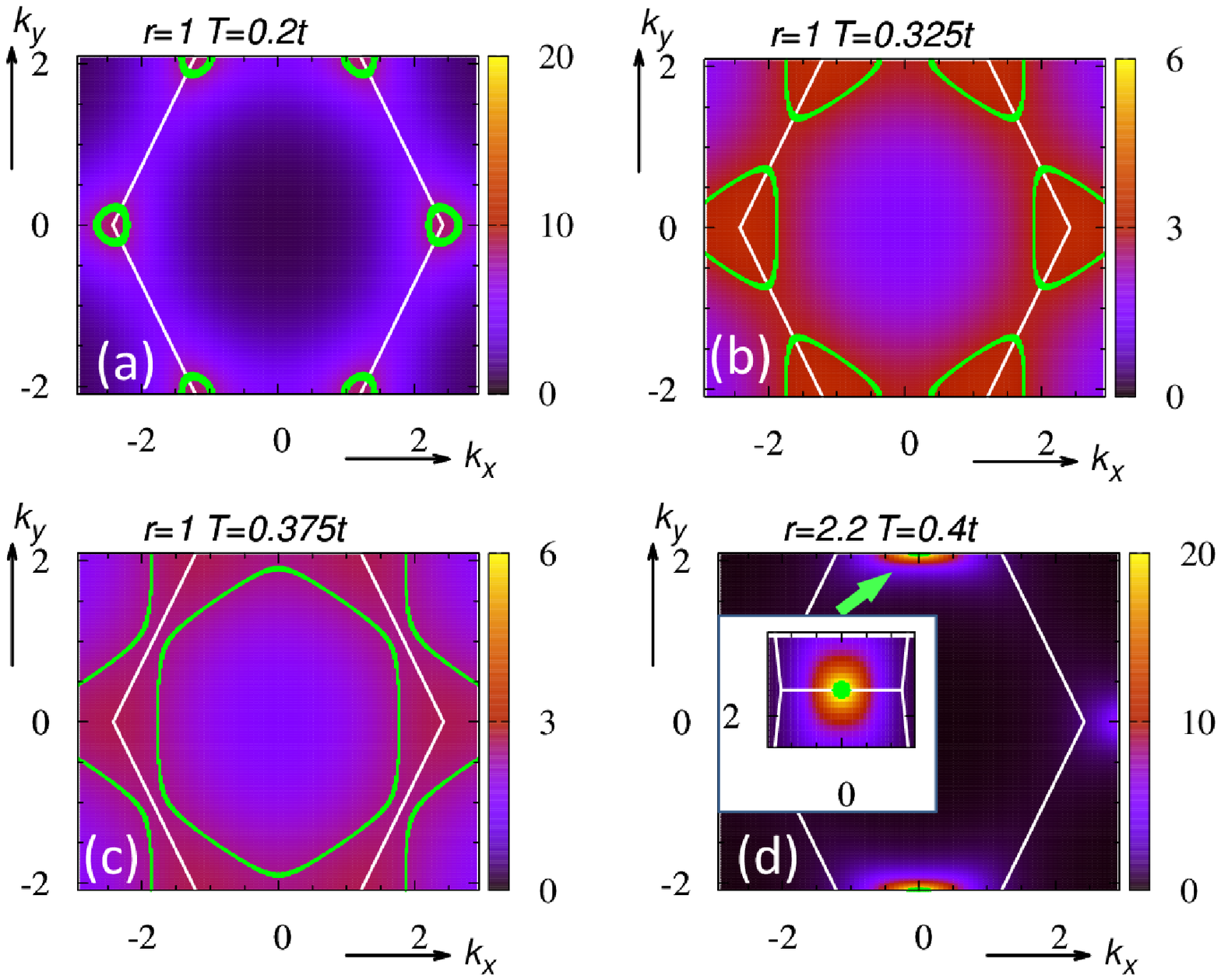}
\end{center}
\end{minipage}
\caption{(Color Online).
Momentum-revolved spectral weight for several values of temperature at $(U_A,U_B)=(10t,5t)$.
White lines illustrate the hexagonal BZ. 
The effective Hamiltonian becomes defective on the green lines.
Panels (a), (b), and (c) are the data for $r=1$ where the system has two Dirac cones, while panel (d) shows the data for $r=2.2$ where the Dirac cones are absent.
The inset of panel (d) shows the magnification of the data around the BZ boundary.
}
\label{fig: Ak}
\end{figure}

Emergence of the SPERs can be observed in the region where the interaction strength is weaker than the critical value of the Mott transition. 
In Fig.~\ref{fig: Ak}, the momentum-resolved spectral weight is plotted for several values of temperature at $(U_A, U_B) = (10t, 5t)$.
The BZ is illustrated by the white hexagon in this figure. 
For $T=0.2t$, we can observe the SPERs indicated by green lines around corners of the BZ [see Fig.~\ref{fig: Ak}(a)], meaning that Dirac cones change into exceptional rings due to the imaginary part of the self-energy.
Increasing temperature enhances the imaginary part of the self-energy. Correspondingly, the SPERs become large. 
Interestingly, we can observe Fermi planes accompanying the SPERs because the energy gap of the $H_{eff}(0,\bm{k})$ is pure imaginary in the region of the BZ enclosed by SPERs [see Fig.~\ref{fig: Ak}(b)]. 
With further increasing temperature, SPERs emerging from two distinct Dirac cones finally merge and change into a single exceptional ring [see Fig.~\ref{fig: Ak}(c)]. 
We note that the presence of Dirac cones for the non-interacting case is not a necessary condition; even in the case for $r=2.2$ where the Dirac cones are gapped, we can still find the emergence of SPERs [see Fig.~\ref{fig: Ak}(d)].

%
%%%%%%%%%%%%%%%%%%%%%%%%%
%%%%%%%%%%%%%%%%%%%%%%%%%
%
%

In the above, we have demonstrated the emergence of SPERs with extended chiral symmetry in the honeycomb lattice model where the single-particle spectrum is governed by the non-Hermitian Dirac Hamiltonian with $l=2$.
In addition, we have numerically observed that Fermi planes accompanying SPERs enhance the specific heat as discussed in Sec.~\ref{sec: cv} of the supplemental material~\cite{suppl}.
We note that the appearance of Fermi planes would serve as a signal of SPERs in cold atom experiments. The measurement of the single-particle spectrum has already been carried out in the context of the pseudo-gap for fermionic cold atoms~\cite{Gaebler_ARPES_cold_NatPhys10,Feld_ARPES_cold_Nat11}.

%%%%%%%%%%%%%%%%%%
\textit{Robustness of SPERs beyond Dirac Hamiltonians.--}
%%%%%%%%%%%%%%%%%%
So far, we have seen the emergence of SPERs when the non-Hermitian effective Hamiltonian is reduced to the Dirac Hamiltonian. Intriguingly, the SPERs with extended chiral symmetry survive even beyond the Dirac Hamiltonian, which is related to a zero-th Chern number.

In order to see the robustness, we first show that a Chern number can be introduced for the following traceless Hamiltonian 
$H'(\bm{k}):=H_{eff}(0,\bm{k})-\left[ \mathrm{tr}H_{eff}(0,\bm{k})/N \right] \1$ with $N:=\mathrm{dim}\ H_{eff}(0,\bm{k})$. 
This matrix also satisfies the extended chiral symmetry because the relation $\mathrm{tr}H_{eff}(0,\bm{k}) \in i\mathbb{R}$ holds.
Here, we expand the Hilbert space and define the following Hermitian Hamiltonian $\tilde{H}(\bm{k})$~\cite{ZPGong_PRX18}:
%%%%%%%%%%%%%%%%%%
\begin{eqnarray}
\tilde{H}(\bm{k})
 &:=& 
 \left(
\begin{array}{cc}
0 & H'(\bm{k}) \\
H'^\dagger(\bm{k}) & 0
\end{array}
\right)_\rho.
\end{eqnarray}
%%%%%%%%%%%%%%%%%%
$\tilde{H}(\bm{k})$ anti-commutes with the following two matrices:
$
\tilde{\Gamma} = U_\Gamma \otimes \rho_1
$ 
and 
$
\tilde{\Sigma}=\1 \otimes \rho_3
$, 
where the Pauli matrices $\rho$'s act on the expanded Hilbert space. In other words, the Hermitian matrix $\tilde{H}(\bm{k})$ satisfies the chiral symmetry both for $\tilde{\Gamma}$ and $\tilde{\Sigma}$.
Thus, we can block-diagonalize the Hamiltonian $\tilde{H}$ with plus and minus sectors of $M:=i\tilde{\Sigma}\tilde{\Gamma}$ $(M^2=\1)$.
Noticing the relations $\{M,\tilde{\Sigma} \}=\{M,\tilde{\Gamma} \}=0$, we can see that the chiral symmetry is not closed for each subsector of the Hilbert space, meaning that the block-diagonalized Hamiltonian for the plus (minus) sector $\tilde{H}_{+(-)}$ belongs to the symmetry class A (no symmetry). 
Therefore, one can define the zero-th Chern number which corresponds to the number of occupied states of $\tilde{H}_{+}(\bm{k})$ at a given point in the BZ. 
We note that $\mathrm{det}H'(\bm{k})$ vanishes between points in the BZ where the zero-th Chern number takes distinct values. This is because $\mathrm{det}\tilde{H}$ vanishes if and only if $\mathrm{det}H'=0$. Since the BZ is two-dimensional, points satisfying $\mathrm{det}H'(\bm{k})=0$ form rings.
Therefore, we end up with the following two scenarios when the zero-th Chern number changes in the BZ. One of them is that exceptional points simply survive even beyond the Dirac Hamiltonian.  The other is that the Hamiltonian $H'(\bm{k})$ becomes diagonalizable and take two eigenstates with zero eigenvalue. In that case, SPERs become open strings.  

Now, we numerically demonstrate the robustness of the SPERs beyond Dirac Hamiltonian for $\mathrm{dim}\ H=4$ by taking the following Hamiltonian as an example.
%%%%%%%%%%%%%%%%%%
\begin{eqnarray}
H_{eff}(0,\bm{k}) &=& b_1(\bm{k}) \tau_1 \sigma_3 +b_2(\bm{k}) \tau_2 \sigma_3 + V\tau_3 \sigma_1 +id_3\tau_3\sigma_3, \nonumber 
\end{eqnarray}
%%%%%%%%%%%%%%%%%%
where $b_1$ and $b_2$ are defined in Eq.~(\ref{eq: bs_honeycomb}). Here $V$ and $d_3$ are real constants. 
The above non-Hermitian Hamiltonian satisfies Eq.~(\ref{eq: chiral_Heff_main}) with $U_{\Gamma}=\tau_3\sigma_3$.
%%%%%%%%%%%%%%%%%%%%%%%%%
\begin{figure}[!h]
\begin{minipage}{0.475\hsize}
\begin{center}
\includegraphics[width=\hsize,clip]{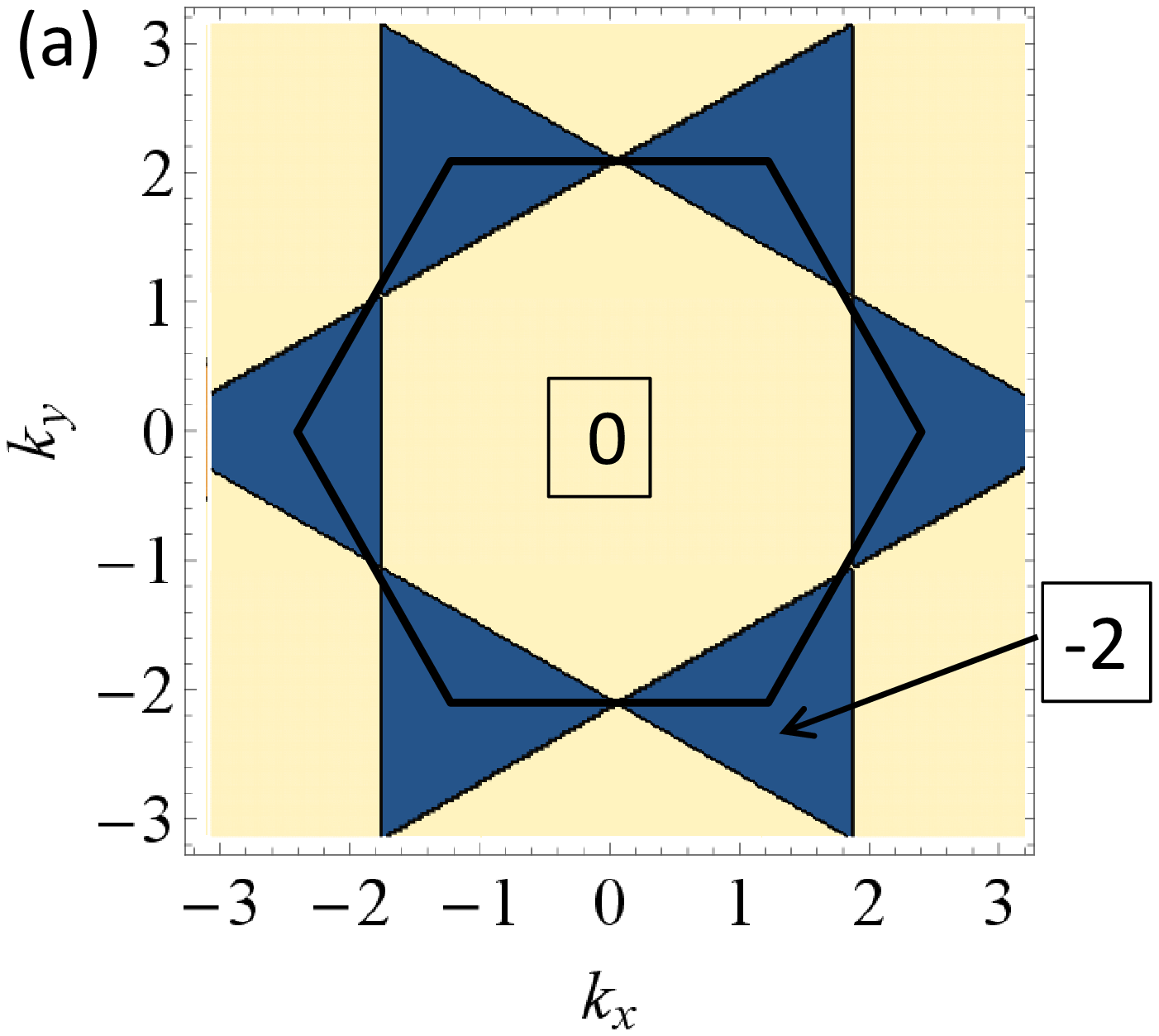}
\end{center}
\end{minipage}
\begin{minipage}{0.475\hsize}
\begin{center}
\includegraphics[width=\hsize,clip]{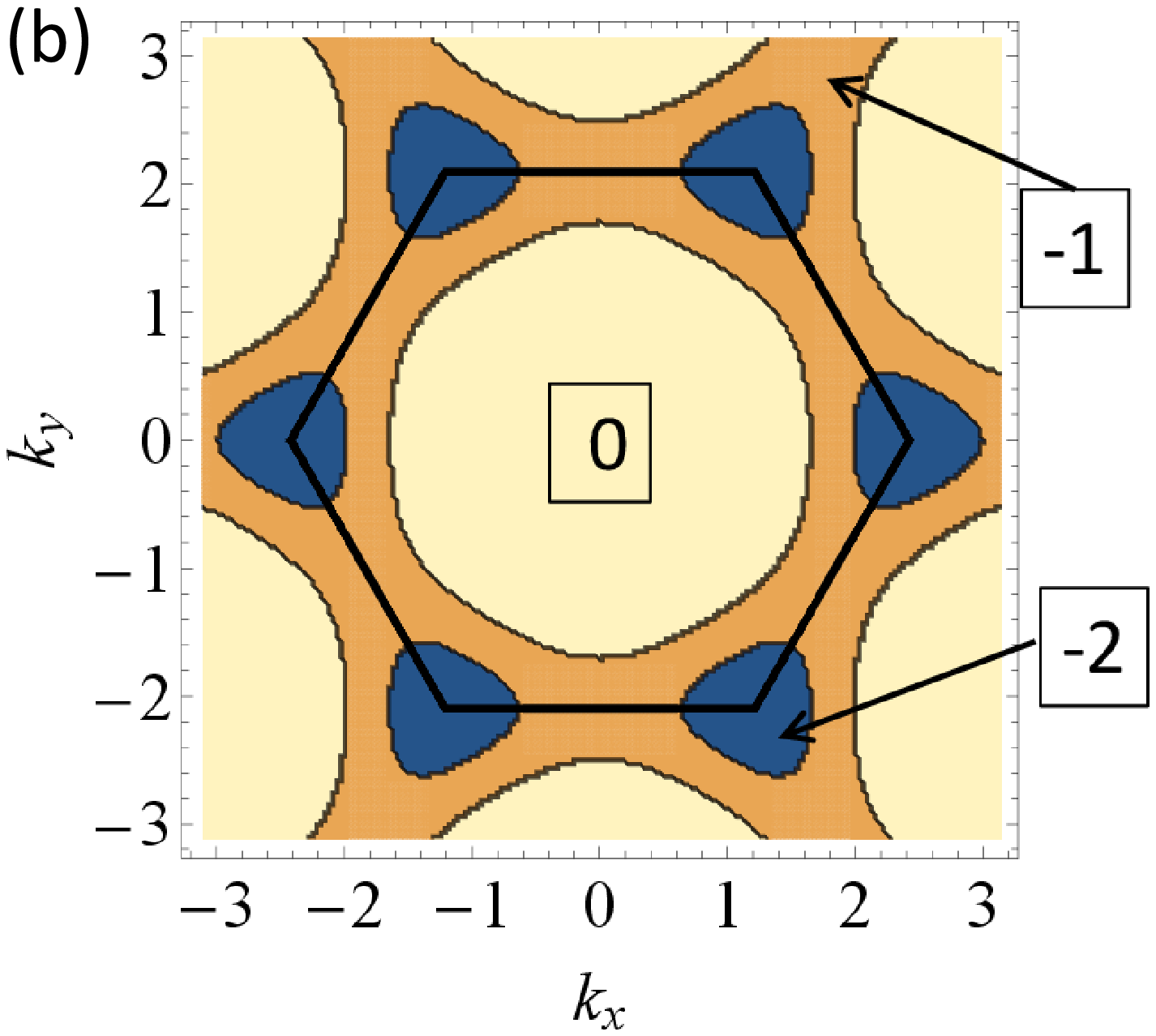}
\end{center}
\end{minipage}
\caption{(Color Online).
Color map of the zero-th Chern number for $t=r=d=1$. 
Panel (a) [(b)] shows the data for $V=0$ [$V=0.3t$]. 
Here, the Chern number at each point is indicated by a number enclosed with a box. 
SPERs are represented with black lines.
The black hexagon illustrates the BZ.
Here, we set zero of the Chern number so that it takes ``0" when the $4\times 4$-matrix $H_+(\bm{k})$ has two occupied states.
}
\label{fig: BeyondDH_Ch}
\end{figure}
%%%%%%%%%%%%%%%%%%%%%%%%%
SPERs and the zero-th Chern number are plotted in Fig.~\ref{fig: BeyondDH_Ch}. 
The energy bands are plotted in Sec.~\ref{sec: BeyondDH_E} of the supplemental material~\cite{suppl}. 
For $V=0$ the problem is reduced to the case of $2\times 2$ matrices which we have discussed previously. 
Correspondingly, each band is doubly degenerate. In this case, the zero-th Chen number for the $4\times 4$ Hamiltonian takes $-2$ or $0$ [see Fig.~\ref{fig: BeyondDH_Ch}(a)]. 
At the boundary the SPERs illustrated by black lines appear. 
Turning on $V$, lifts the degeneracy of each band, and the zero-th Chern number takes $-2$, $-1$, or $0$ [see Fig.~\ref{fig: BeyondDH_Ch}(b)]. Even in this case, we can observe exceptional rings illustrated by black lines, demonstrating that robustness of SPERs is related by the zero-th Chern number.

%%%%%%%%%%%%%%%%%%%%%%%%%%%%%%%%%%%%
\textit{Summary.--}
%%%%%%%%%%%%%%%%%%%%%%%%%%%%%%%%%%%%
In order to elucidate the impacts of symmetry on exceptional points, we have analyzed the strongly correlated systems with chrial symmetry in equilibrium. 
Intriguingly, we have revealed the emergence of the novel topological degeneracies, the SPERs, which provides a new direction for analysis of exceptional points. 
Specifically, we have discovered the SPER based on the following results. 
By focusing on the non-Hermitian Dirac Hamiltonian, we have uncovered that extended chiral symmetry~(\ref{eq: chiral_Heff_main}) results in the SPERs in the two-dimensional BZ. 
Notably, the SPER in two dimensions is beyond the characterization of exceptional points with vorticity.
In addition, by applying the DMFT+NRG, the emergence of SPERs has been demonstrated for the correlated honeycomb lattice. The SPERs in the honeycomb lattice are considered to be observed for cold atoms.
Intriguingly, SPERs are robust even beyond the Dirac model; SPER emerges in the region of the BZ where the zero-th Chern number changes. 

We finish this letter with a comment on the following open issue to be addressed: extending the symmetry protection of non-Hermitian topological degeneracies to other symmetry classes of the ten-fold way~\cite{Schnyder_classification_free_2008,Kitaev_classification_free_2009,Ryu_classification_free_2010}. Namely, the classification of topological insulators/superconductors has revealed that time-reversal, particle-hole, and chiral symmetry serve as primitive local symmetry protecting the topology of insulators/superconductors. 
Therefore, it is natural to expect that systems with time-reversal/particle-hole symmetry host new symmetry-protected topological degeneracies. We leave this question as our future work to be addressed.

%%%%%%%%%%%%%%%%%%%%%%%%%%%%%%%%%%%%
\textit{Note added.--}
%%%%%%%%%%%%%%%%%%%%%%%%%%%%%%%%%%%%
While finising this paper, we noticed the paper~[\onlinecite{Budich_SPER_arXiv18}] appeared on arXiv which has some overlap with our results.

%%%%%%%%%%%%%%%%%%%%%%%%%%%%%%%%%%%%
\textit{
Acknowledgements.--
}
%%%%%%%%%%%%%%%%%%%%%%%%%%%%%%%%%%%%
This work is partly supported by JSPS KAKENHI Grants No.~JP15H05855, No.~JP16K05501, No.~JP16K13845, No.~JP17H06138, No.~JP18H01140, No.~JP18H04316, No.~JP18K03511, and No.~JP18H05842.
The numerical calculations were performed on supercomputer at the ISSP 
in the University of Tokyo, and the SR16000 at YITP in Kyoto University.

%\bibliography{Ering}

\begin{thebibliography}{66}%
\makeatletter
\providecommand \@ifxundefined [1]{%
 \@ifx{#1\undefined}
}%
\providecommand \@ifnum [1]{%
 \ifnum #1\expandafter \@firstoftwo
 \else \expandafter \@secondoftwo
 \fi
}%
\providecommand \@ifx [1]{%
 \ifx #1\expandafter \@firstoftwo
 \else \expandafter \@secondoftwo
 \fi
}%
\providecommand \natexlab [1]{#1}%
\providecommand \enquote  [1]{``#1''}%
\providecommand \bibnamefont  [1]{#1}%
\providecommand \bibfnamefont [1]{#1}%
\providecommand \citenamefont [1]{#1}%
\providecommand \href@noop [0]{\@secondoftwo}%
\providecommand \href [0]{\begingroup \@sanitize@url \@href}%
\providecommand \@href[1]{\@@startlink{#1}\@@href}%
\providecommand \@@href[1]{\endgroup#1\@@endlink}%
\providecommand \@sanitize@url [0]{\catcode `\\12\catcode `\$12\catcode
  `\&12\catcode `\#12\catcode `\^12\catcode `\_12\catcode `\%12\relax}%
\providecommand \@@startlink[1]{}%
\providecommand \@@endlink[0]{}%
\providecommand \url  [0]{\begingroup\@sanitize@url \@url }%
\providecommand \@url [1]{\endgroup\@href {#1}{\urlprefix }}%
\providecommand \urlprefix  [0]{URL }%
\providecommand \Eprint [0]{\href }%
\providecommand \doibase [0]{http://dx.doi.org/}%
\providecommand \selectlanguage [0]{\@gobble}%
\providecommand \bibinfo  [0]{\@secondoftwo}%
\providecommand \bibfield  [0]{\@secondoftwo}%
\providecommand \translation [1]{[#1]}%
\providecommand \BibitemOpen [0]{}%
\providecommand \bibitemStop [0]{}%
\providecommand \bibitemNoStop [0]{.\EOS\space}%
\providecommand \EOS [0]{\spacefactor3000\relax}%
\providecommand \BibitemShut  [1]{\csname bibitem#1\endcsname}%
\let\auto@bib@innerbib\@empty
%</preamble>
\bibitem [{\citenamefont {Hasan}\ and\ \citenamefont
  {Kane}(2010)}]{TI_review_Hasan10}%
  \BibitemOpen
  \bibfield  {author} {\bibinfo {author} {\bibfnamefont {M.~Z.}\ \bibnamefont
  {Hasan}}\ and\ \bibinfo {author} {\bibfnamefont {C.~L.}\ \bibnamefont
  {Kane}},\ }\href {\doibase 10.1103/RevModPhys.82.3045} {\bibfield  {journal}
  {\bibinfo  {journal} {Rev. Mod. Phys.}\ }\textbf {\bibinfo {volume} {82}},\
  \bibinfo {pages} {3045} (\bibinfo {year} {2010})}\BibitemShut {NoStop}%
\bibitem [{\citenamefont {Qi}\ and\ \citenamefont
  {Zhang}(2011)}]{TI_review_Qi10}%
  \BibitemOpen
  \bibfield  {author} {\bibinfo {author} {\bibfnamefont {X.-L.}\ \bibnamefont
  {Qi}}\ and\ \bibinfo {author} {\bibfnamefont {S.-C.}\ \bibnamefont {Zhang}},\
  }\href {\doibase 10.1103/RevModPhys.83.1057} {\bibfield  {journal} {\bibinfo
  {journal} {Rev. Mod. Phys.}\ }\textbf {\bibinfo {volume} {83}},\ \bibinfo
  {pages} {1057} (\bibinfo {year} {2011})}\BibitemShut {NoStop}%
\bibitem [{\citenamefont {Ando}(2013)}]{Ando_review_JPSJ13}%
  \BibitemOpen
  \bibfield  {author} {\bibinfo {author} {\bibfnamefont {Y.}~\bibnamefont
  {Ando}},\ }\href {\doibase 10.7566/JPSJ.82.102001} {\bibfield  {journal}
  {\bibinfo  {journal} {Journal of the Physical Society of Japan}\ }\textbf
  {\bibinfo {volume} {82}},\ \bibinfo {pages} {102001} (\bibinfo {year}
  {2013})},\ \Eprint
  {http://arxiv.org/abs/https://doi.org/10.7566/JPSJ.82.102001}
  {https://doi.org/10.7566/JPSJ.82.102001} \BibitemShut {NoStop}%
\bibitem [{\citenamefont {Weng}\ \emph {et~al.}(2015)\citenamefont {Weng},
  \citenamefont {Fang}, \citenamefont {Fang}, \citenamefont {Bernevig},\ and\
  \citenamefont {Dai}}]{Weng_TaAs_LDA_PRX15}%
  \BibitemOpen
  \bibfield  {author} {\bibinfo {author} {\bibfnamefont {H.}~\bibnamefont
  {Weng}}, \bibinfo {author} {\bibfnamefont {C.}~\bibnamefont {Fang}}, \bibinfo
  {author} {\bibfnamefont {Z.}~\bibnamefont {Fang}}, \bibinfo {author}
  {\bibfnamefont {B.~A.}\ \bibnamefont {Bernevig}}, \ and\ \bibinfo {author}
  {\bibfnamefont {X.}~\bibnamefont {Dai}},\ }\href {\doibase
  10.1103/PhysRevX.5.011029} {\bibfield  {journal} {\bibinfo  {journal} {Phys.
  Rev. X}\ }\textbf {\bibinfo {volume} {5}},\ \bibinfo {pages} {011029}
  (\bibinfo {year} {2015})}\BibitemShut {NoStop}%
\bibitem [{\citenamefont {Xu}\ \emph {et~al.}(2015)\citenamefont {Xu},
  \citenamefont {Belopolski}, \citenamefont {Alidoust}, \citenamefont
  {Neupane}, \citenamefont {Bian}, \citenamefont {Zhang}, \citenamefont
  {Sankar}, \citenamefont {Chang}, \citenamefont {Yuan}, \citenamefont {Lee},
  \citenamefont {Huang}, \citenamefont {Zheng}, \citenamefont {Ma},
  \citenamefont {Sanchez}, \citenamefont {Wang}, \citenamefont {Bansil},
  \citenamefont {Chou}, \citenamefont {Shibayev}, \citenamefont {Lin},
  \citenamefont {Jia},\ and\ \citenamefont {Hasan}}]{Xu_TaAs_arpes_Sci15}%
  \BibitemOpen
  \bibfield  {author} {\bibinfo {author} {\bibfnamefont {S.-Y.}\ \bibnamefont
  {Xu}}, \bibinfo {author} {\bibfnamefont {I.}~\bibnamefont {Belopolski}},
  \bibinfo {author} {\bibfnamefont {N.}~\bibnamefont {Alidoust}}, \bibinfo
  {author} {\bibfnamefont {M.}~\bibnamefont {Neupane}}, \bibinfo {author}
  {\bibfnamefont {G.}~\bibnamefont {Bian}}, \bibinfo {author} {\bibfnamefont
  {C.}~\bibnamefont {Zhang}}, \bibinfo {author} {\bibfnamefont
  {R.}~\bibnamefont {Sankar}}, \bibinfo {author} {\bibfnamefont
  {G.}~\bibnamefont {Chang}}, \bibinfo {author} {\bibfnamefont
  {Z.}~\bibnamefont {Yuan}}, \bibinfo {author} {\bibfnamefont {C.-C.}\
  \bibnamefont {Lee}}, \bibinfo {author} {\bibfnamefont {S.-M.}\ \bibnamefont
  {Huang}}, \bibinfo {author} {\bibfnamefont {H.}~\bibnamefont {Zheng}},
  \bibinfo {author} {\bibfnamefont {J.}~\bibnamefont {Ma}}, \bibinfo {author}
  {\bibfnamefont {D.~S.}\ \bibnamefont {Sanchez}}, \bibinfo {author}
  {\bibfnamefont {B.}~\bibnamefont {Wang}}, \bibinfo {author} {\bibfnamefont
  {A.}~\bibnamefont {Bansil}}, \bibinfo {author} {\bibfnamefont
  {F.}~\bibnamefont {Chou}}, \bibinfo {author} {\bibfnamefont {P.~P.}\
  \bibnamefont {Shibayev}}, \bibinfo {author} {\bibfnamefont {H.}~\bibnamefont
  {Lin}}, \bibinfo {author} {\bibfnamefont {S.}~\bibnamefont {Jia}}, \ and\
  \bibinfo {author} {\bibfnamefont {M.~Z.}\ \bibnamefont {Hasan}},\ }\href
  {\doibase 10.1126/science.aaa9297} {\bibfield  {journal} {\bibinfo  {journal}
  {Science}\ }\textbf {\bibinfo {volume} {349}},\ \bibinfo {pages} {613}
  (\bibinfo {year} {2015})},\ \Eprint
  {http://arxiv.org/abs/http://science.sciencemag.org/content/349/6248/613.full.pdf}
  {http://science.sciencemag.org/content/349/6248/613.full.pdf} \BibitemShut
  {NoStop}%
\bibitem [{\citenamefont {Wan}\ \emph {et~al.}(2011)\citenamefont {Wan},
  \citenamefont {Turner}, \citenamefont {Vishwanath},\ and\ \citenamefont
  {Savrasov}}]{Weyl_theor_PRB11}%
  \BibitemOpen
  \bibfield  {author} {\bibinfo {author} {\bibfnamefont {X.}~\bibnamefont
  {Wan}}, \bibinfo {author} {\bibfnamefont {A.~M.}\ \bibnamefont {Turner}},
  \bibinfo {author} {\bibfnamefont {A.}~\bibnamefont {Vishwanath}}, \ and\
  \bibinfo {author} {\bibfnamefont {S.~Y.}\ \bibnamefont {Savrasov}},\ }\href
  {\doibase 10.1103/PhysRevB.83.205101} {\bibfield  {journal} {\bibinfo
  {journal} {Phys. Rev. B}\ }\textbf {\bibinfo {volume} {83}},\ \bibinfo
  {pages} {205101} (\bibinfo {year} {2011})}\BibitemShut {NoStop}%
\bibitem [{\citenamefont {Burkov}\ and\ \citenamefont
  {Balents}(2011)}]{Burkov_Weyl_theor_PRL11}%
  \BibitemOpen
  \bibfield  {author} {\bibinfo {author} {\bibfnamefont {A.~A.}\ \bibnamefont
  {Burkov}}\ and\ \bibinfo {author} {\bibfnamefont {L.}~\bibnamefont
  {Balents}},\ }\href {\doibase 10.1103/PhysRevLett.107.127205} {\bibfield
  {journal} {\bibinfo  {journal} {Phys. Rev. Lett.}\ }\textbf {\bibinfo
  {volume} {107}},\ \bibinfo {pages} {127205} (\bibinfo {year}
  {2011})}\BibitemShut {NoStop}%
\bibitem [{\citenamefont {Kane}\ and\ \citenamefont
  {Mele}(2005{\natexlab{a}})}]{Kane_PRL05_1}%
  \BibitemOpen
  \bibfield  {author} {\bibinfo {author} {\bibfnamefont {C.~L.}\ \bibnamefont
  {Kane}}\ and\ \bibinfo {author} {\bibfnamefont {E.~J.}\ \bibnamefont
  {Mele}},\ }\href {\doibase 10.1103/PhysRevLett.95.146802} {\bibfield
  {journal} {\bibinfo  {journal} {Phys. Rev. Lett.}\ }\textbf {\bibinfo
  {volume} {95}},\ \bibinfo {pages} {146802} (\bibinfo {year}
  {2005}{\natexlab{a}})}\BibitemShut {NoStop}%
\bibitem [{\citenamefont {Kane}\ and\ \citenamefont
  {Mele}(2005{\natexlab{b}})}]{Kane_PRL05_2}%
  \BibitemOpen
  \bibfield  {author} {\bibinfo {author} {\bibfnamefont {C.~L.}\ \bibnamefont
  {Kane}}\ and\ \bibinfo {author} {\bibfnamefont {E.~J.}\ \bibnamefont
  {Mele}},\ }\href {\doibase 10.1103/PhysRevLett.95.226801} {\bibfield
  {journal} {\bibinfo  {journal} {Phys. Rev. Lett.}\ }\textbf {\bibinfo
  {volume} {95}},\ \bibinfo {pages} {226801} (\bibinfo {year}
  {2005}{\natexlab{b}})}\BibitemShut {NoStop}%
\bibitem [{\citenamefont {Bernevig}\ \emph {et~al.}(2006)\citenamefont
  {Bernevig}, \citenamefont {Hughes},\ and\ \citenamefont
  {Zhang}}]{Bernevig_QSH_Science06}%
  \BibitemOpen
  \bibfield  {author} {\bibinfo {author} {\bibfnamefont {B.~A.}\ \bibnamefont
  {Bernevig}}, \bibinfo {author} {\bibfnamefont {T.~L.}\ \bibnamefont
  {Hughes}}, \ and\ \bibinfo {author} {\bibfnamefont {S.-C.}\ \bibnamefont
  {Zhang}},\ }\href {\doibase 10.1126/science.1133734} {\bibfield  {journal}
  {\bibinfo  {journal} {Science}\ }\textbf {\bibinfo {volume} {314}},\ \bibinfo
  {pages} {1757} (\bibinfo {year} {2006})},\ \Eprint
  {http://arxiv.org/abs/http://science.sciencemag.org/content/314/5806/1757.full.pdf}
  {http://science.sciencemag.org/content/314/5806/1757.full.pdf} \BibitemShut
  {NoStop}%
\bibitem [{\citenamefont {K{\"o}nig}\ \emph {et~al.}(2007)\citenamefont
  {K{\"o}nig}, \citenamefont {Wiedmann}, \citenamefont {Br{\"u}ne},
  \citenamefont {Roth}, \citenamefont {Buhmann}, \citenamefont {Molenkamp},
  \citenamefont {Qi},\ and\ \citenamefont {Zhang}}]{Konig_Science07}%
  \BibitemOpen
  \bibfield  {author} {\bibinfo {author} {\bibfnamefont {M.}~\bibnamefont
  {K{\"o}nig}}, \bibinfo {author} {\bibfnamefont {S.}~\bibnamefont {Wiedmann}},
  \bibinfo {author} {\bibfnamefont {C.}~\bibnamefont {Br{\"u}ne}}, \bibinfo
  {author} {\bibfnamefont {A.}~\bibnamefont {Roth}}, \bibinfo {author}
  {\bibfnamefont {H.}~\bibnamefont {Buhmann}}, \bibinfo {author} {\bibfnamefont
  {L.~W.}\ \bibnamefont {Molenkamp}}, \bibinfo {author} {\bibfnamefont {X.-L.}\
  \bibnamefont {Qi}}, \ and\ \bibinfo {author} {\bibfnamefont {S.-C.}\
  \bibnamefont {Zhang}},\ }\href {\doibase 10.1126/science.1148047} {\bibfield
  {journal} {\bibinfo  {journal} {Science}\ }\textbf {\bibinfo {volume}
  {318}},\ \bibinfo {pages} {766} (\bibinfo {year} {2007})},\ \Eprint
  {http://arxiv.org/abs/http://science.sciencemag.org/content/318/5851/766.full.pdf}
  {http://science.sciencemag.org/content/318/5851/766.full.pdf} \BibitemShut
  {NoStop}%
\bibitem [{\citenamefont {Kitaev}(2001)}]{Kitaev_chain_01}%
  \BibitemOpen
  \bibfield  {author} {\bibinfo {author} {\bibfnamefont {A.~Y.}\ \bibnamefont
  {Kitaev}},\ }\href {http://stacks.iop.org/1063-7869/44/i=10S/a=S29}
  {\bibfield  {journal} {\bibinfo  {journal} {Physics-Uspekhi}\ }\textbf
  {\bibinfo {volume} {44}},\ \bibinfo {pages} {131} (\bibinfo {year}
  {2001})}\BibitemShut {NoStop}%
\bibitem [{\citenamefont {Ryu}\ and\ \citenamefont
  {Hatsugai}(2002)}]{Majorana_Ryu_RPL02}%
  \BibitemOpen
  \bibfield  {author} {\bibinfo {author} {\bibfnamefont {S.}~\bibnamefont
  {Ryu}}\ and\ \bibinfo {author} {\bibfnamefont {Y.}~\bibnamefont {Hatsugai}},\
  }\href {\doibase 10.1103/PhysRevLett.89.077002} {\bibfield  {journal}
  {\bibinfo  {journal} {Phys. Rev. Lett.}\ }\textbf {\bibinfo {volume} {89}},\
  \bibinfo {pages} {077002} (\bibinfo {year} {2002})}\BibitemShut {NoStop}%
\bibitem [{\citenamefont {Alicea}(2012)}]{Alicia_Majorana_review12}%
  \BibitemOpen
  \bibfield  {author} {\bibinfo {author} {\bibfnamefont {J.}~\bibnamefont
  {Alicea}},\ }\href {http://stacks.iop.org/0034-4885/75/i=7/a=076501}
  {\bibfield  {journal} {\bibinfo  {journal} {Reports on Progress in Physics}\
  }\textbf {\bibinfo {volume} {75}},\ \bibinfo {pages} {076501} (\bibinfo
  {year} {2012})}\BibitemShut {NoStop}%
\bibitem [{\citenamefont {Mourik}\ \emph {et~al.}(2012)\citenamefont {Mourik},
  \citenamefont {Zuo}, \citenamefont {Frolov}, \citenamefont {Plissard},
  \citenamefont {Bakkers},\ and\ \citenamefont
  {Kouwenhoven}}]{Majorana_Mourik}%
  \BibitemOpen
  \bibfield  {author} {\bibinfo {author} {\bibfnamefont {V.}~\bibnamefont
  {Mourik}}, \bibinfo {author} {\bibfnamefont {K.}~\bibnamefont {Zuo}},
  \bibinfo {author} {\bibfnamefont {S.~M.}\ \bibnamefont {Frolov}}, \bibinfo
  {author} {\bibfnamefont {S.~R.}\ \bibnamefont {Plissard}}, \bibinfo {author}
  {\bibfnamefont {E.~P. A.~M.}\ \bibnamefont {Bakkers}}, \ and\ \bibinfo
  {author} {\bibfnamefont {L.~P.}\ \bibnamefont {Kouwenhoven}},\ }\href
  {\doibase 10.1126/science.1222360} {\bibfield  {journal} {\bibinfo  {journal}
  {Science}\ }\textbf {\bibinfo {volume} {336}},\ \bibinfo {pages} {1003}
  (\bibinfo {year} {2012})}\BibitemShut {NoStop}%
\bibitem [{\citenamefont {Rokhinson}\ \emph {et~al.}(2012)\citenamefont
  {Rokhinson}, \citenamefont {Liu},\ and\ \citenamefont
  {Furdyna}}]{Majorana_Rokhinson2012}%
  \BibitemOpen
  \bibfield  {author} {\bibinfo {author} {\bibfnamefont {L.~P.}\ \bibnamefont
  {Rokhinson}}, \bibinfo {author} {\bibfnamefont {X.}~\bibnamefont {Liu}}, \
  and\ \bibinfo {author} {\bibfnamefont {J.~K.}\ \bibnamefont {Furdyna}},\
  }\href@noop {} {\bibfield  {journal} {\bibinfo  {journal} {Nature Physics}\
  }\textbf {\bibinfo {volume} {8}},\ \bibinfo {pages} {795} (\bibinfo {year}
  {2012})}\BibitemShut {NoStop}%
\bibitem [{\citenamefont {Das}\ \emph {et~al.}(2012)\citenamefont {Das},
  \citenamefont {Ronen}, \citenamefont {Most}, \citenamefont {Oreg},
  \citenamefont {Heiblum},\ and\ \citenamefont {Shtrikman}}]{Majorana_Das2012}%
  \BibitemOpen
  \bibfield  {author} {\bibinfo {author} {\bibfnamefont {A.}~\bibnamefont
  {Das}}, \bibinfo {author} {\bibfnamefont {Y.}~\bibnamefont {Ronen}}, \bibinfo
  {author} {\bibfnamefont {Y.}~\bibnamefont {Most}}, \bibinfo {author}
  {\bibfnamefont {Y.}~\bibnamefont {Oreg}}, \bibinfo {author} {\bibfnamefont
  {M.}~\bibnamefont {Heiblum}}, \ and\ \bibinfo {author} {\bibfnamefont
  {H.}~\bibnamefont {Shtrikman}},\ }\href@noop {} {\bibfield  {journal}
  {\bibinfo  {journal} {Nature Physics}\ }\textbf {\bibinfo {volume} {8}},\
  \bibinfo {pages} {887} (\bibinfo {year} {2012})}\BibitemShut {NoStop}%
\bibitem [{\citenamefont {Sato}\ and\ \citenamefont
  {Fujimoto}(2016)}]{Sato_JPSJ16}%
  \BibitemOpen
  \bibfield  {author} {\bibinfo {author} {\bibfnamefont {M.}~\bibnamefont
  {Sato}}\ and\ \bibinfo {author} {\bibfnamefont {S.}~\bibnamefont
  {Fujimoto}},\ }\href {\doibase 10.7566/JPSJ.85.072001} {\bibfield  {journal}
  {\bibinfo  {journal} {Journal of the Physical Society of Japan}\ }\textbf
  {\bibinfo {volume} {85}},\ \bibinfo {pages} {072001} (\bibinfo {year}
  {2016})},\ \Eprint
  {http://arxiv.org/abs/https://doi.org/10.7566/JPSJ.85.072001}
  {https://doi.org/10.7566/JPSJ.85.072001} \BibitemShut {NoStop}%
\bibitem [{\citenamefont {Morimoto}\ and\ \citenamefont
  {Furusaki}(2014)}]{Morimoto_Weyl_RPB14}%
  \BibitemOpen
  \bibfield  {author} {\bibinfo {author} {\bibfnamefont {T.}~\bibnamefont
  {Morimoto}}\ and\ \bibinfo {author} {\bibfnamefont {A.}~\bibnamefont
  {Furusaki}},\ }\href {\doibase 10.1103/PhysRevB.89.235127} {\bibfield
  {journal} {\bibinfo  {journal} {Phys. Rev. B}\ }\textbf {\bibinfo {volume}
  {89}},\ \bibinfo {pages} {235127} (\bibinfo {year} {2014})}\BibitemShut
  {NoStop}%
\bibitem [{\citenamefont {Zhao}\ \emph {et~al.}(2016)\citenamefont {Zhao},
  \citenamefont {Schnyder},\ and\ \citenamefont {Wang}}]{Zhao_PRL16}%
  \BibitemOpen
  \bibfield  {author} {\bibinfo {author} {\bibfnamefont {Y.~X.}\ \bibnamefont
  {Zhao}}, \bibinfo {author} {\bibfnamefont {A.~P.}\ \bibnamefont {Schnyder}},
  \ and\ \bibinfo {author} {\bibfnamefont {Z.~D.}\ \bibnamefont {Wang}},\
  }\href {\doibase 10.1103/PhysRevLett.116.156402} {\bibfield  {journal}
  {\bibinfo  {journal} {Phys. Rev. Lett.}\ }\textbf {\bibinfo {volume} {116}},\
  \bibinfo {pages} {156402} (\bibinfo {year} {2016})}\BibitemShut {NoStop}%
\bibitem [{\citenamefont {Hatano}\ and\ \citenamefont
  {Nelson}(1996)}]{Hatano-NelsonPRL96}%
  \BibitemOpen
  \bibfield  {author} {\bibinfo {author} {\bibfnamefont {N.}~\bibnamefont
  {Hatano}}\ and\ \bibinfo {author} {\bibfnamefont {D.~R.}\ \bibnamefont
  {Nelson}},\ }\href {\doibase 10.1103/PhysRevLett.77.570} {\bibfield
  {journal} {\bibinfo  {journal} {Phys. Rev. Lett.}\ }\textbf {\bibinfo
  {volume} {77}},\ \bibinfo {pages} {570} (\bibinfo {year} {1996})}\BibitemShut
  {NoStop}%
\bibitem [{\citenamefont {Fukui}\ and\ \citenamefont
  {Kawakami}(1998)}]{Fukui-KawakamiPRB98}%
  \BibitemOpen
  \bibfield  {author} {\bibinfo {author} {\bibfnamefont {T.}~\bibnamefont
  {Fukui}}\ and\ \bibinfo {author} {\bibfnamefont {N.}~\bibnamefont
  {Kawakami}},\ }\href {\doibase 10.1103/PhysRevB.58.16051} {\bibfield
  {journal} {\bibinfo  {journal} {Phys. Rev. B}\ }\textbf {\bibinfo {volume}
  {58}},\ \bibinfo {pages} {16051} (\bibinfo {year} {1998})}\BibitemShut
  {NoStop}%
\bibitem [{\citenamefont {Bender}\ and\ \citenamefont
  {Boettcher}(1998)}]{CMBender_PRL98}%
  \BibitemOpen
  \bibfield  {author} {\bibinfo {author} {\bibfnamefont {C.~M.}\ \bibnamefont
  {Bender}}\ and\ \bibinfo {author} {\bibfnamefont {S.}~\bibnamefont
  {Boettcher}},\ }\href {\doibase 10.1103/PhysRevLett.80.5243} {\bibfield
  {journal} {\bibinfo  {journal} {Phys. Rev. Lett.}\ }\textbf {\bibinfo
  {volume} {80}},\ \bibinfo {pages} {5243} (\bibinfo {year}
  {1998})}\BibitemShut {NoStop}%
\bibitem [{\citenamefont {Guo}\ \emph {et~al.}(2009)\citenamefont {Guo},
  \citenamefont {Salamo}, \citenamefont {Duchesne}, \citenamefont {Morandotti},
  \citenamefont {Volatier-Ravat}, \citenamefont {Aimez}, \citenamefont
  {Siviloglou},\ and\ \citenamefont {Christodoulides}}]{Guo_nHExp_PRL09}%
  \BibitemOpen
  \bibfield  {author} {\bibinfo {author} {\bibfnamefont {A.}~\bibnamefont
  {Guo}}, \bibinfo {author} {\bibfnamefont {G.~J.}\ \bibnamefont {Salamo}},
  \bibinfo {author} {\bibfnamefont {D.}~\bibnamefont {Duchesne}}, \bibinfo
  {author} {\bibfnamefont {R.}~\bibnamefont {Morandotti}}, \bibinfo {author}
  {\bibfnamefont {M.}~\bibnamefont {Volatier-Ravat}}, \bibinfo {author}
  {\bibfnamefont {V.}~\bibnamefont {Aimez}}, \bibinfo {author} {\bibfnamefont
  {G.~A.}\ \bibnamefont {Siviloglou}}, \ and\ \bibinfo {author} {\bibfnamefont
  {D.~N.}\ \bibnamefont {Christodoulides}},\ }\href {\doibase
  10.1103/PhysRevLett.103.093902} {\bibfield  {journal} {\bibinfo  {journal}
  {Phys. Rev. Lett.}\ }\textbf {\bibinfo {volume} {103}},\ \bibinfo {pages}
  {093902} (\bibinfo {year} {2009})}\BibitemShut {NoStop}%
\bibitem [{\citenamefont {R{\"u}ter}\ \emph {et~al.}(2010)\citenamefont
  {R{\"u}ter}, \citenamefont {Makris}, \citenamefont {El-Ganainy},
  \citenamefont {Christodoulides}, \citenamefont {Segev},\ and\ \citenamefont
  {Kip}}]{Ruter_nHExp_NatPhys10}%
  \BibitemOpen
  \bibfield  {author} {\bibinfo {author} {\bibfnamefont {C.~E.}\ \bibnamefont
  {R{\"u}ter}}, \bibinfo {author} {\bibfnamefont {K.~G.}\ \bibnamefont
  {Makris}}, \bibinfo {author} {\bibfnamefont {R.}~\bibnamefont {El-Ganainy}},
  \bibinfo {author} {\bibfnamefont {D.~N.}\ \bibnamefont {Christodoulides}},
  \bibinfo {author} {\bibfnamefont {M.}~\bibnamefont {Segev}}, \ and\ \bibinfo
  {author} {\bibfnamefont {D.}~\bibnamefont {Kip}},\ }\href@noop {} {\bibfield
  {journal} {\bibinfo  {journal} {Nature physics}\ }\textbf {\bibinfo {volume}
  {6}},\ \bibinfo {pages} {192} (\bibinfo {year} {2010})}\BibitemShut {NoStop}%
\bibitem [{\citenamefont {Regensburger}\ \emph {et~al.}(2012)\citenamefont
  {Regensburger}, \citenamefont {Bersch}, \citenamefont {Miri}, \citenamefont
  {Onishchukov}, \citenamefont {Christodoulides},\ and\ \citenamefont
  {Peschel}}]{Regensburger_nHExp_Nat12}%
  \BibitemOpen
  \bibfield  {author} {\bibinfo {author} {\bibfnamefont {A.}~\bibnamefont
  {Regensburger}}, \bibinfo {author} {\bibfnamefont {C.}~\bibnamefont
  {Bersch}}, \bibinfo {author} {\bibfnamefont {M.-A.}\ \bibnamefont {Miri}},
  \bibinfo {author} {\bibfnamefont {G.}~\bibnamefont {Onishchukov}}, \bibinfo
  {author} {\bibfnamefont {D.~N.}\ \bibnamefont {Christodoulides}}, \ and\
  \bibinfo {author} {\bibfnamefont {U.}~\bibnamefont {Peschel}},\ }\href@noop
  {} {\bibfield  {journal} {\bibinfo  {journal} {Nature}\ }\textbf {\bibinfo
  {volume} {488}},\ \bibinfo {pages} {167} (\bibinfo {year}
  {2012})}\BibitemShut {NoStop}%
\bibitem [{\citenamefont {Ashida}\ \emph {et~al.}(2016)\citenamefont {Ashida},
  \citenamefont {Furukawa},\ and\ \citenamefont {Ueda}}]{Ashida_PRA16}%
  \BibitemOpen
  \bibfield  {author} {\bibinfo {author} {\bibfnamefont {Y.}~\bibnamefont
  {Ashida}}, \bibinfo {author} {\bibfnamefont {S.}~\bibnamefont {Furukawa}}, \
  and\ \bibinfo {author} {\bibfnamefont {M.}~\bibnamefont {Ueda}},\ }\href
  {\doibase 10.1103/PhysRevA.94.053615} {\bibfield  {journal} {\bibinfo
  {journal} {Phys. Rev. A}\ }\textbf {\bibinfo {volume} {94}},\ \bibinfo
  {pages} {053615} (\bibinfo {year} {2016})}\BibitemShut {NoStop}%
\bibitem [{\citenamefont {Ashida}\ \emph {et~al.}(2017)\citenamefont {Ashida},
  \citenamefont {Furukawa},\ and\ \citenamefont {Ueda}}]{YAshida_NatCom17}%
  \BibitemOpen
  \bibfield  {author} {\bibinfo {author} {\bibfnamefont {Y.}~\bibnamefont
  {Ashida}}, \bibinfo {author} {\bibfnamefont {S.}~\bibnamefont {Furukawa}}, \
  and\ \bibinfo {author} {\bibfnamefont {M.}~\bibnamefont {Ueda}},\ }\href
  {http://dx.doi.org/10.1038/ncomms15791} {\bibfield  {journal} {\bibinfo
  {journal} {Nature Communications}\ }\textbf {\bibinfo {volume} {8}},\
  \bibinfo {pages} {15791 EP } (\bibinfo {year} {2017})},\ \bibinfo {note}
  {article}\BibitemShut {NoStop}%
\bibitem [{\citenamefont {Kozii}\ and\ \citenamefont
  {Fu}(2017)}]{VKozii2017_non-Hermi}%
  \BibitemOpen
  \bibfield  {author} {\bibinfo {author} {\bibfnamefont {V.}~\bibnamefont
  {Kozii}}\ and\ \bibinfo {author} {\bibfnamefont {L.}~\bibnamefont {Fu}},\
  }\href@noop {} {\bibfield  {journal} {\bibinfo  {journal} {arXiv preprint
  arXiv:1708.05841}\ } (\bibinfo {year} {2017})}\BibitemShut {NoStop}%
\bibitem [{\citenamefont {Zhen}\ \emph {et~al.}(2015)\citenamefont {Zhen},
  \citenamefont {Hsu}, \citenamefont {Igarashi}, \citenamefont {Lu},
  \citenamefont {Kaminer}, \citenamefont {Pick}, \citenamefont {Chua},
  \citenamefont {Joannopoulos},\ and\ \citenamefont
  {Soljacic}}]{Zhen_EP_Nature15}%
  \BibitemOpen
  \bibfield  {author} {\bibinfo {author} {\bibfnamefont {B.}~\bibnamefont
  {Zhen}}, \bibinfo {author} {\bibfnamefont {C.~W.}\ \bibnamefont {Hsu}},
  \bibinfo {author} {\bibfnamefont {Y.}~\bibnamefont {Igarashi}}, \bibinfo
  {author} {\bibfnamefont {L.}~\bibnamefont {Lu}}, \bibinfo {author}
  {\bibfnamefont {I.}~\bibnamefont {Kaminer}}, \bibinfo {author} {\bibfnamefont
  {A.}~\bibnamefont {Pick}}, \bibinfo {author} {\bibfnamefont {S.-L.}\
  \bibnamefont {Chua}}, \bibinfo {author} {\bibfnamefont {J.~D.}\ \bibnamefont
  {Joannopoulos}}, \ and\ \bibinfo {author} {\bibfnamefont {M.}~\bibnamefont
  {Soljacic}},\ }\href {http://dx.doi.org/10.1038/nature14889} {\bibfield
  {journal} {\bibinfo  {journal} {Nature}\ }\textbf {\bibinfo {volume} {525}},\
  \bibinfo {pages} {354 EP } (\bibinfo {year} {2015})}\BibitemShut {NoStop}%
\bibitem [{\citenamefont {Shen}\ \emph {et~al.}(2018)\citenamefont {Shen},
  \citenamefont {Zhen},\ and\ \citenamefont {Fu}}]{HShen2017_non-Hermi}%
  \BibitemOpen
  \bibfield  {author} {\bibinfo {author} {\bibfnamefont {H.}~\bibnamefont
  {Shen}}, \bibinfo {author} {\bibfnamefont {B.}~\bibnamefont {Zhen}}, \ and\
  \bibinfo {author} {\bibfnamefont {L.}~\bibnamefont {Fu}},\ }\href {\doibase
  10.1103/PhysRevLett.120.146402} {\bibfield  {journal} {\bibinfo  {journal}
  {Phys. Rev. Lett.}\ }\textbf {\bibinfo {volume} {120}},\ \bibinfo {pages}
  {146402} (\bibinfo {year} {2018})}\BibitemShut {NoStop}%
\bibitem [{\citenamefont {Xu}\ \emph {et~al.}(2017)\citenamefont {Xu},
  \citenamefont {Wang},\ and\ \citenamefont
  {Duan}}]{YXuPRL17_exceptional_ring}%
  \BibitemOpen
  \bibfield  {author} {\bibinfo {author} {\bibfnamefont {Y.}~\bibnamefont
  {Xu}}, \bibinfo {author} {\bibfnamefont {S.-T.}\ \bibnamefont {Wang}}, \ and\
  \bibinfo {author} {\bibfnamefont {L.-M.}\ \bibnamefont {Duan}},\ }\href
  {\doibase 10.1103/PhysRevLett.118.045701} {\bibfield  {journal} {\bibinfo
  {journal} {Phys. Rev. Lett.}\ }\textbf {\bibinfo {volume} {118}},\ \bibinfo
  {pages} {045701} (\bibinfo {year} {2017})}\BibitemShut {NoStop}%
\bibitem [{\citenamefont {Yoshida}\ \emph
  {et~al.}(2018{\natexlab{a}})\citenamefont {Yoshida}, \citenamefont {Peters},\
  and\ \citenamefont {Kawakami}}]{Yoshida_nHKondo_PRB18}%
  \BibitemOpen
  \bibfield  {author} {\bibinfo {author} {\bibfnamefont {T.}~\bibnamefont
  {Yoshida}}, \bibinfo {author} {\bibfnamefont {R.}~\bibnamefont {Peters}}, \
  and\ \bibinfo {author} {\bibfnamefont {N.}~\bibnamefont {Kawakami}},\ }\href
  {\doibase 10.1103/PhysRevB.98.035141} {\bibfield  {journal} {\bibinfo
  {journal} {Phys. Rev. B}\ }\textbf {\bibinfo {volume} {98}},\ \bibinfo
  {pages} {035141} (\bibinfo {year} {2018}{\natexlab{a}})}\BibitemShut
  {NoStop}%
\bibitem [{\citenamefont {Esaki}\ \emph {et~al.}(2011)\citenamefont {Esaki},
  \citenamefont {Sato}, \citenamefont {Hasebe},\ and\ \citenamefont
  {Kohmoto}}]{EsakiSatoHasebeKhomoto_12}%
  \BibitemOpen
  \bibfield  {author} {\bibinfo {author} {\bibfnamefont {K.}~\bibnamefont
  {Esaki}}, \bibinfo {author} {\bibfnamefont {M.}~\bibnamefont {Sato}},
  \bibinfo {author} {\bibfnamefont {K.}~\bibnamefont {Hasebe}}, \ and\ \bibinfo
  {author} {\bibfnamefont {M.}~\bibnamefont {Kohmoto}},\ }\href {\doibase
  10.1103/PhysRevB.84.205128} {\bibfield  {journal} {\bibinfo  {journal} {Phys.
  Rev. B}\ }\textbf {\bibinfo {volume} {84}},\ \bibinfo {pages} {205128}
  (\bibinfo {year} {2011})}\BibitemShut {NoStop}%
\bibitem [{\citenamefont {Sato}\ \emph {et~al.}()\citenamefont {Sato},
  \citenamefont {Hasebe}, \citenamefont {Esaki},\ and\ \citenamefont
  {Kohmoto}}]{SatoEsakiHasebeKhomoto_12}%
  \BibitemOpen
  \bibfield  {author} {\bibinfo {author} {\bibfnamefont {M.}~\bibnamefont
  {Sato}}, \bibinfo {author} {\bibfnamefont {K.}~\bibnamefont {Hasebe}},
  \bibinfo {author} {\bibfnamefont {K.}~\bibnamefont {Esaki}}, \ and\ \bibinfo
  {author} {\bibfnamefont {M.}~\bibnamefont {Kohmoto}},\ }\href@noop {}
  {\bibfield  {journal} {\bibinfo  {journal} {Progress of Theoretical Physics}\
  }\textbf {\bibinfo {volume} {127}},\ \bibinfo {pages} {937}}\BibitemShut
  {NoStop}%
\bibitem [{\citenamefont {San-Jose}\ \emph {et~al.}(2016)\citenamefont
  {San-Jose}, \citenamefont {Cayao}, \citenamefont {Prada},\ and\ \citenamefont
  {Aguado}}]{San-Jose2016}%
  \BibitemOpen
  \bibfield  {author} {\bibinfo {author} {\bibfnamefont {P.}~\bibnamefont
  {San-Jose}}, \bibinfo {author} {\bibfnamefont {J.}~\bibnamefont {Cayao}},
  \bibinfo {author} {\bibfnamefont {E.}~\bibnamefont {Prada}}, \ and\ \bibinfo
  {author} {\bibfnamefont {R.}~\bibnamefont {Aguado}},\ }\href
  {https://doi.org/10.1038/srep21427} {\bibfield  {journal} {\bibinfo
  {journal} {Scientific Reports}\ }\textbf {\bibinfo {volume} {6}},\ \bibinfo
  {pages} {21427 EP } (\bibinfo {year} {2016})},\ \bibinfo {note}
  {article}\BibitemShut {NoStop}%
\bibitem [{\citenamefont {Lee}(2016)}]{TELeePRL16_Half_quantized}%
  \BibitemOpen
  \bibfield  {author} {\bibinfo {author} {\bibfnamefont {T.~E.}\ \bibnamefont
  {Lee}},\ }\href {\doibase 10.1103/PhysRevLett.116.133903} {\bibfield
  {journal} {\bibinfo  {journal} {Phys. Rev. Lett.}\ }\textbf {\bibinfo
  {volume} {116}},\ \bibinfo {pages} {133903} (\bibinfo {year}
  {2016})}\BibitemShut {NoStop}%
\bibitem [{\citenamefont {Gong}\ \emph {et~al.}(2017)\citenamefont {Gong},
  \citenamefont {Higashikawa},\ and\ \citenamefont {Ueda}}]{ZPGong_PRL17}%
  \BibitemOpen
  \bibfield  {author} {\bibinfo {author} {\bibfnamefont {Z.}~\bibnamefont
  {Gong}}, \bibinfo {author} {\bibfnamefont {S.}~\bibnamefont {Higashikawa}}, \
  and\ \bibinfo {author} {\bibfnamefont {M.}~\bibnamefont {Ueda}},\ }\href
  {\doibase 10.1103/PhysRevLett.118.200401} {\bibfield  {journal} {\bibinfo
  {journal} {Phys. Rev. Lett.}\ }\textbf {\bibinfo {volume} {118}},\ \bibinfo
  {pages} {200401} (\bibinfo {year} {2017})}\BibitemShut {NoStop}%
\bibitem [{\citenamefont {Zyuzin}\ and\ \citenamefont
  {Zyuzin}(2018)}]{Zyuzin_disorder_WeylPRB18}%
  \BibitemOpen
  \bibfield  {author} {\bibinfo {author} {\bibfnamefont {A.~A.}\ \bibnamefont
  {Zyuzin}}\ and\ \bibinfo {author} {\bibfnamefont {A.~Y.}\ \bibnamefont
  {Zyuzin}},\ }\href {\doibase 10.1103/PhysRevB.97.041203} {\bibfield
  {journal} {\bibinfo  {journal} {Phys. Rev. B}\ }\textbf {\bibinfo {volume}
  {97}},\ \bibinfo {pages} {041203} (\bibinfo {year} {2018})}\BibitemShut
  {NoStop}%
\bibitem [{\citenamefont {Gong}\ \emph {et~al.}(2018)\citenamefont {Gong},
  \citenamefont {Ashida}, \citenamefont {Kawabata}, \citenamefont {Takasan},
  \citenamefont {Higashikawa},\ and\ \citenamefont {Ueda}}]{ZPGong_PRX18}%
  \BibitemOpen
  \bibfield  {author} {\bibinfo {author} {\bibfnamefont {Z.}~\bibnamefont
  {Gong}}, \bibinfo {author} {\bibfnamefont {Y.}~\bibnamefont {Ashida}},
  \bibinfo {author} {\bibfnamefont {K.}~\bibnamefont {Kawabata}}, \bibinfo
  {author} {\bibfnamefont {K.}~\bibnamefont {Takasan}}, \bibinfo {author}
  {\bibfnamefont {S.}~\bibnamefont {Higashikawa}}, \ and\ \bibinfo {author}
  {\bibfnamefont {M.}~\bibnamefont {Ueda}},\ }\href {\doibase
  10.1103/PhysRevX.8.031079} {\bibfield  {journal} {\bibinfo  {journal} {Phys.
  Rev. X}\ }\textbf {\bibinfo {volume} {8}},\ \bibinfo {pages} {031079}
  (\bibinfo {year} {2018})}\BibitemShut {NoStop}%
\bibitem [{\citenamefont {Kawabata}\ \emph {et~al.}(2018)\citenamefont
  {Kawabata}, \citenamefont {Ashida}, \citenamefont {Katsura},\ and\
  \citenamefont {Ueda}}]{KKawabata_arXiv18}%
  \BibitemOpen
  \bibfield  {author} {\bibinfo {author} {\bibfnamefont {K.}~\bibnamefont
  {Kawabata}}, \bibinfo {author} {\bibfnamefont {Y.}~\bibnamefont {Ashida}},
  \bibinfo {author} {\bibfnamefont {H.}~\bibnamefont {Katsura}}, \ and\
  \bibinfo {author} {\bibfnamefont {M.}~\bibnamefont {Ueda}},\ }\href@noop {}
  {\bibfield  {journal} {\bibinfo  {journal} {arXiv preprint arXiv:1801.00499}\
  } (\bibinfo {year} {2018})}\BibitemShut {NoStop}%
\bibitem [{\citenamefont {Yao}\ and\ \citenamefont
  {Wang}(2018)}]{Yao_1DBBC_PRL2018}%
  \BibitemOpen
  \bibfield  {author} {\bibinfo {author} {\bibfnamefont {S.}~\bibnamefont
  {Yao}}\ and\ \bibinfo {author} {\bibfnamefont {Z.}~\bibnamefont
  {Wang}},\ }\href {\doibase 10.1103/PhysRevLett.121.086803} {\bibfield
  {journal} {\bibinfo  {journal} {Phys. Rev. Lett.}\ }\textbf {\bibinfo
  {volume} {121}},\ \bibinfo {pages} {086803} (\bibinfo {year}
  {2018})}\BibitemShut {NoStop}%
\bibitem [{\citenamefont {Yao}\ \emph {et~al.}(2018)\citenamefont
  {Yao}, \citenamefont {Song},\ and\ \citenamefont
  {Wang}}]{Yao_2DBBC_PRL2018}%
  \BibitemOpen
  \bibfield  {author} {\bibinfo {author} {\bibfnamefont {S.}~\bibnamefont
  {Yao}}, \bibinfo {author} {\bibfnamefont {F.}~\bibnamefont {Song}}, \
  and\ \bibinfo {author} {\bibfnamefont {Z.}~\bibnamefont {Wang}},\ }\href
  {\doibase 10.1103/PhysRevLett.121.136802} {\bibfield  {journal} {\bibinfo
  {journal} {Phys. Rev. Lett.}\ }\textbf {\bibinfo {volume} {121}},\ \bibinfo
  {pages} {136802} (\bibinfo {year} {2018})}\BibitemShut {NoStop}%
\bibitem [{\citenamefont {Metzner}\ and\ \citenamefont
  {Vollhardt}(1989)}]{WMetznerPRL89_DMFT}%
  \BibitemOpen
  \bibfield  {author} {\bibinfo {author} {\bibfnamefont {W.}~\bibnamefont
  {Metzner}}\ and\ \bibinfo {author} {\bibfnamefont {D.}~\bibnamefont
  {Vollhardt}},\ }\href {\doibase 10.1103/PhysRevLett.62.324} {\bibfield
  {journal} {\bibinfo  {journal} {Phys. Rev. Lett.}\ }\textbf {\bibinfo
  {volume} {62}},\ \bibinfo {pages} {324} (\bibinfo {year} {1989})}\BibitemShut
  {NoStop}%
\bibitem [{\citenamefont {M{\"u}ller-Hartmann}(1989)}]{MHartmannZP89_DMFT}%
  \BibitemOpen
  \bibfield  {author} {\bibinfo {author} {\bibfnamefont {E.}~\bibnamefont
  {M{\"u}ller-Hartmann}},\ }\href {\doibase 10.1007/BF01311397} {\bibfield
  {journal} {\bibinfo  {journal} {Zeitschrift f{\"u}r Physik B Condensed
  Matter}\ }\textbf {\bibinfo {volume} {74}},\ \bibinfo {pages} {507} (\bibinfo
  {year} {1989})}\BibitemShut {NoStop}%
\bibitem [{\citenamefont {Georges}\ \emph {et~al.}(1996)\citenamefont
  {Georges}, \citenamefont {Kotliar}, \citenamefont {Krauth},\ and\
  \citenamefont {Rozenberg}}]{AGeorgesRMP96_DMFT}%
  \BibitemOpen
  \bibfield  {author} {\bibinfo {author} {\bibfnamefont {A.}~\bibnamefont
  {Georges}}, \bibinfo {author} {\bibfnamefont {G.}~\bibnamefont {Kotliar}},
  \bibinfo {author} {\bibfnamefont {W.}~\bibnamefont {Krauth}}, \ and\ \bibinfo
  {author} {\bibfnamefont {M.~J.}\ \bibnamefont {Rozenberg}},\ }\href {\doibase
  10.1103/RevModPhys.68.13} {\bibfield  {journal} {\bibinfo  {journal} {Rev.
  Mod. Phys.}\ }\textbf {\bibinfo {volume} {68}},\ \bibinfo {pages} {13}
  (\bibinfo {year} {1996})}\BibitemShut {NoStop}%
\bibitem [{\citenamefont {Wilson}(1975)}]{KWilsonRMP75_NRG}%
  \BibitemOpen
  \bibfield  {author} {\bibinfo {author} {\bibfnamefont {K.~G.}\ \bibnamefont
  {Wilson}},\ }\href {\doibase 10.1103/RevModPhys.47.773} {\bibfield  {journal}
  {\bibinfo  {journal} {Rev. Mod. Phys.}\ }\textbf {\bibinfo {volume} {47}},\
  \bibinfo {pages} {773} (\bibinfo {year} {1975})}\BibitemShut {NoStop}%
\bibitem [{\citenamefont {Peters}\ \emph {et~al.}(2006)\citenamefont {Peters},
  \citenamefont {Pruschke},\ and\ \citenamefont {Anders}}]{RPetersPRB06_NRG}%
  \BibitemOpen
  \bibfield  {author} {\bibinfo {author} {\bibfnamefont {R.}~\bibnamefont
  {Peters}}, \bibinfo {author} {\bibfnamefont {T.}~\bibnamefont {Pruschke}}, \
  and\ \bibinfo {author} {\bibfnamefont {F.~B.}\ \bibnamefont {Anders}},\
  }\href {\doibase 10.1103/PhysRevB.74.245114} {\bibfield  {journal} {\bibinfo
  {journal} {Phys. Rev. B}\ }\textbf {\bibinfo {volume} {74}},\ \bibinfo
  {pages} {245114} (\bibinfo {year} {2006})}\BibitemShut {NoStop}%
\bibitem [{\citenamefont {Bulla}\ \emph {et~al.}(2008)\citenamefont {Bulla},
  \citenamefont {Costi},\ and\ \citenamefont {Pruschke}}]{RBullaRMP08_NRG}%
  \BibitemOpen
  \bibfield  {author} {\bibinfo {author} {\bibfnamefont {R.}~\bibnamefont
  {Bulla}}, \bibinfo {author} {\bibfnamefont {T.~A.}\ \bibnamefont {Costi}}, \
  and\ \bibinfo {author} {\bibfnamefont {T.}~\bibnamefont {Pruschke}},\ }\href
  {\doibase 10.1103/RevModPhys.80.395} {\bibfield  {journal} {\bibinfo
  {journal} {Rev. Mod. Phys.}\ }\textbf {\bibinfo {volume} {80}},\ \bibinfo
  {pages} {395} (\bibinfo {year} {2008})}\BibitemShut {NoStop}%
\bibitem [{\citenamefont {Hatsugai}(2006)}]{Hatsugai_mb_chiral_JPSJ06}%
  \BibitemOpen
  \bibfield  {author} {\bibinfo {author} {\bibfnamefont {Y.}~\bibnamefont
  {Hatsugai}},\ }\href {\doibase 10.1143/JPSJ.75.123601} {\bibfield  {journal}
  {\bibinfo  {journal} {Journal of the Physical Society of Japan}\ }\textbf
  {\bibinfo {volume} {75}},\ \bibinfo {pages} {123601} (\bibinfo {year}
  {2006})},\ \Eprint
  {http://arxiv.org/abs/https://doi.org/10.1143/JPSJ.75.123601}
  {https://doi.org/10.1143/JPSJ.75.123601} \BibitemShut {NoStop}%
\bibitem [{\citenamefont {Gurarie}(2011)}]{Gurarie_PRB11}%
  \BibitemOpen
  \bibfield  {author} {\bibinfo {author} {\bibfnamefont {V.}~\bibnamefont
  {Gurarie}},\ }\href {\doibase 10.1103/PhysRevB.83.085426} {\bibfield
  {journal} {\bibinfo  {journal} {Phys. Rev. B}\ }\textbf {\bibinfo {volume}
  {83}},\ \bibinfo {pages} {085426} (\bibinfo {year} {2011})}\BibitemShut
  {NoStop}%
\bibitem [{sup()}]{suppl}%
  \BibitemOpen
  \href@noop {} {}\bibinfo {note} {{ See Supplemental Material. }}\BibitemShut
  {NoStop}%
\bibitem [{\citenamefont {Becker}\ \emph {et~al.}(2010)\citenamefont {Becker},
  \citenamefont {Soltan-Panahi}, \citenamefont {Kronj{\"a}ger}, \citenamefont
  {D{\"o}rscher}, \citenamefont {Bongs},\ and\ \citenamefont
  {Sengstock}}]{Becker_honey_coldIOP10}%
  \BibitemOpen
  \bibfield  {author} {\bibinfo {author} {\bibfnamefont {C.}~\bibnamefont
  {Becker}}, \bibinfo {author} {\bibfnamefont {P.}~\bibnamefont
  {Soltan-Panahi}}, \bibinfo {author} {\bibfnamefont {J.}~\bibnamefont
  {Kronj{\"a}ger}}, \bibinfo {author} {\bibfnamefont {S.}~\bibnamefont
  {D{\"o}rscher}}, \bibinfo {author} {\bibfnamefont {K.}~\bibnamefont {Bongs}},
  \ and\ \bibinfo {author} {\bibfnamefont {K.}~\bibnamefont {Sengstock}},\
  }\href {http://stacks.iop.org/1367-2630/12/i=6/a=065025} {\bibfield
  {journal} {\bibinfo  {journal} {New Journal of Physics}\ }\textbf {\bibinfo
  {volume} {12}},\ \bibinfo {pages} {065025} (\bibinfo {year}
  {2010})}\BibitemShut {NoStop}%
\bibitem [{\citenamefont {Yamazaki}\ \emph {et~al.}(2010)\citenamefont
  {Yamazaki}, \citenamefont {Taie}, \citenamefont {Sugawa},\ and\ \citenamefont
  {Takahashi}}]{RYamazaki_PRL10}%
  \BibitemOpen
  \bibfield  {author} {\bibinfo {author} {\bibfnamefont {R.}~\bibnamefont
  {Yamazaki}}, \bibinfo {author} {\bibfnamefont {S.}~\bibnamefont {Taie}},
  \bibinfo {author} {\bibfnamefont {S.}~\bibnamefont {Sugawa}}, \ and\ \bibinfo
  {author} {\bibfnamefont {Y.}~\bibnamefont {Takahashi}},\ }\href {\doibase
  10.1103/PhysRevLett.105.050405} {\bibfield  {journal} {\bibinfo  {journal}
  {Phys. Rev. Lett.}\ }\textbf {\bibinfo {volume} {105}},\ \bibinfo {pages}
  {050405} (\bibinfo {year} {2010})}\BibitemShut {NoStop}%
\bibitem [{\citenamefont {Clark}\ \emph {et~al.}(2015)\citenamefont {Clark},
  \citenamefont {Ha}, \citenamefont {Xu},\ and\ \citenamefont
  {Chin}}]{LWClark_PRL15}%
  \BibitemOpen
  \bibfield  {author} {\bibinfo {author} {\bibfnamefont {L.~W.}\ \bibnamefont
  {Clark}}, \bibinfo {author} {\bibfnamefont {L.-C.}\ \bibnamefont {Ha}},
  \bibinfo {author} {\bibfnamefont {C.-Y.}\ \bibnamefont {Xu}}, \ and\ \bibinfo
  {author} {\bibfnamefont {C.}~\bibnamefont {Chin}},\ }\href {\doibase
  10.1103/PhysRevLett.115.155301} {\bibfield  {journal} {\bibinfo  {journal}
  {Phys. Rev. Lett.}\ }\textbf {\bibinfo {volume} {115}},\ \bibinfo {pages}
  {155301} (\bibinfo {year} {2015})}\BibitemShut {NoStop}%
\bibitem [{\citenamefont {Manmana}\ \emph {et~al.}(2012)\citenamefont
  {Manmana}, \citenamefont {Essin}, \citenamefont {Noack},\ and\ \citenamefont
  {Gurarie}}]{Manmana_PRB12}%
  \BibitemOpen
  \bibfield  {author} {\bibinfo {author} {\bibfnamefont {S.~R.}\ \bibnamefont
  {Manmana}}, \bibinfo {author} {\bibfnamefont {A.~M.}\ \bibnamefont {Essin}},
  \bibinfo {author} {\bibfnamefont {R.~M.}\ \bibnamefont {Noack}}, \ and\
  \bibinfo {author} {\bibfnamefont {V.}~\bibnamefont {Gurarie}},\ }\href
  {\doibase 10.1103/PhysRevB.86.205119} {\bibfield  {journal} {\bibinfo
  {journal} {Phys. Rev. B}\ }\textbf {\bibinfo {volume} {86}},\ \bibinfo
  {pages} {205119} (\bibinfo {year} {2012})}\BibitemShut {NoStop}%
\bibitem [{\citenamefont {Yoshida}\ \emph {et~al.}(2014)\citenamefont
  {Yoshida}, \citenamefont {Peters}, \citenamefont {Fujimoto},\ and\
  \citenamefont {Kawakami}}]{TMI_YoshidaPRL14}%
  \BibitemOpen
  \bibfield  {author} {\bibinfo {author} {\bibfnamefont {T.}~\bibnamefont
  {Yoshida}}, \bibinfo {author} {\bibfnamefont {R.}~\bibnamefont {Peters}},
  \bibinfo {author} {\bibfnamefont {S.}~\bibnamefont {Fujimoto}}, \ and\
  \bibinfo {author} {\bibfnamefont {N.}~\bibnamefont {Kawakami}},\ }\href
  {\doibase 10.1103/PhysRevLett.112.196404} {\bibfield  {journal} {\bibinfo
  {journal} {Phys. Rev. Lett.}\ }\textbf {\bibinfo {volume} {112}},\ \bibinfo
  {pages} {196404} (\bibinfo {year} {2014})}\BibitemShut {NoStop}%
\bibitem [{\citenamefont {Yoshida}\ \emph
  {et~al.}(2018{\natexlab{b}})\citenamefont {Yoshida}, \citenamefont
  {Danshita}, \citenamefont {Peters},\ and\ \citenamefont
  {Kawakami}}]{Yoshida_ZtoZ4_PRL18}%
  \BibitemOpen
  \bibfield  {author} {\bibinfo {author} {\bibfnamefont {T.}~\bibnamefont
  {Yoshida}}, \bibinfo {author} {\bibfnamefont {I.}~\bibnamefont {Danshita}},
  \bibinfo {author} {\bibfnamefont {R.}~\bibnamefont {Peters}}, \ and\ \bibinfo
  {author} {\bibfnamefont {N.}~\bibnamefont {Kawakami}},\ }\href {\doibase
  10.1103/PhysRevLett.121.025301} {\bibfield  {journal} {\bibinfo  {journal}
  {Phys. Rev. Lett.}\ }\textbf {\bibinfo {volume} {121}},\ \bibinfo {pages}
  {025301} (\bibinfo {year} {2018}{\natexlab{b}})}\BibitemShut {NoStop}%
%
%
%
\bibitem [{foo()}]{footnote_bDMFT}%
  \BibitemOpen
  \href@noop {} {}\bibinfo {note} {{ 
   $\mathrm{Im}\Sigma(i\delta,\bm{k})$ remains diagonal even beyond the DMFT because extended chiral symmetry imposes the following condition: $\tau_3 \mathrm{Im}\Sigma(i\delta,\bm{k}) \tau_3=\mathrm{Im}\Sigma(i\delta,\bm{k})$
  }}\BibitemShut {NoStop}%
%
%
%
\bibitem [{\citenamefont {Yoshida}\ \emph {et~al.}(2012)\citenamefont
  {Yoshida}, \citenamefont {Fujimoto},\ and\ \citenamefont
  {Kawakami}}]{Yoshida_DMFT_PRB12}%
  \BibitemOpen
  \bibfield  {author} {\bibinfo {author} {\bibfnamefont {T.}~\bibnamefont
  {Yoshida}}, \bibinfo {author} {\bibfnamefont {S.}~\bibnamefont {Fujimoto}}, \
  and\ \bibinfo {author} {\bibfnamefont {N.}~\bibnamefont {Kawakami}},\ }\href
  {\doibase 10.1103/PhysRevB.85.125113} {\bibfield  {journal} {\bibinfo
  {journal} {Phys. Rev. B}\ }\textbf {\bibinfo {volume} {85}},\ \bibinfo
  {pages} {125113} (\bibinfo {year} {2012})}\BibitemShut {NoStop}%
\bibitem [{\citenamefont {Yoshida}\ \emph {et~al.}(2013)\citenamefont
  {Yoshida}, \citenamefont {Peters}, \citenamefont {Fujimoto},\ and\
  \citenamefont {Kawakami}}]{Yoshida_PRB13_AFTI}%
  \BibitemOpen
  \bibfield  {author} {\bibinfo {author} {\bibfnamefont {T.}~\bibnamefont
  {Yoshida}}, \bibinfo {author} {\bibfnamefont {R.}~\bibnamefont {Peters}},
  \bibinfo {author} {\bibfnamefont {S.}~\bibnamefont {Fujimoto}}, \ and\
  \bibinfo {author} {\bibfnamefont {N.}~\bibnamefont {Kawakami}},\ }\href
  {\doibase 10.1103/PhysRevB.87.085134} {\bibfield  {journal} {\bibinfo
  {journal} {Phys. Rev. B}\ }\textbf {\bibinfo {volume} {87}},\ \bibinfo
  {pages} {085134} (\bibinfo {year} {2013})}\BibitemShut {NoStop}%
\bibitem [{\citenamefont {Gaebler}\ \emph {et~al.}(2010)\citenamefont
  {Gaebler}, \citenamefont {Stewart}, \citenamefont {Drake}, \citenamefont
  {Jin}, \citenamefont {Perali}, \citenamefont {Pieri},\ and\ \citenamefont
  {Strinati}}]{Gaebler_ARPES_cold_NatPhys10}%
  \BibitemOpen
  \bibfield  {author} {\bibinfo {author} {\bibfnamefont {J.}~\bibnamefont
  {Gaebler}}, \bibinfo {author} {\bibfnamefont {J.}~\bibnamefont {Stewart}},
  \bibinfo {author} {\bibfnamefont {T.}~\bibnamefont {Drake}}, \bibinfo
  {author} {\bibfnamefont {D.}~\bibnamefont {Jin}}, \bibinfo {author}
  {\bibfnamefont {A.}~\bibnamefont {Perali}}, \bibinfo {author} {\bibfnamefont
  {P.}~\bibnamefont {Pieri}}, \ and\ \bibinfo {author} {\bibfnamefont
  {G.}~\bibnamefont {Strinati}},\ }\href@noop {} {\bibfield  {journal}
  {\bibinfo  {journal} {Nature Physics}\ }\textbf {\bibinfo {volume} {6}},\
  \bibinfo {pages} {569} (\bibinfo {year} {2010})}\BibitemShut {NoStop}%
\bibitem [{\citenamefont {Feld}\ \emph {et~al.}(2011)\citenamefont {Feld},
  \citenamefont {Fr{\"o}hlich}, \citenamefont {Vogt}, \citenamefont
  {Koschorreck},\ and\ \citenamefont {K{\"o}hl}}]{Feld_ARPES_cold_Nat11}%
  \BibitemOpen
  \bibfield  {author} {\bibinfo {author} {\bibfnamefont {M.}~\bibnamefont
  {Feld}}, \bibinfo {author} {\bibfnamefont {B.}~\bibnamefont {Fr{\"o}hlich}},
  \bibinfo {author} {\bibfnamefont {E.}~\bibnamefont {Vogt}}, \bibinfo {author}
  {\bibfnamefont {M.}~\bibnamefont {Koschorreck}}, \ and\ \bibinfo {author}
  {\bibfnamefont {M.}~\bibnamefont {K{\"o}hl}},\ }\href@noop {} {\bibfield
  {journal} {\bibinfo  {journal} {Nature}\ }\textbf {\bibinfo {volume} {480}},\
  \bibinfo {pages} {75} (\bibinfo {year} {2011})}\BibitemShut {NoStop}%
\bibitem [{\citenamefont {Schnyder}\ \emph {et~al.}(2008)\citenamefont
  {Schnyder}, \citenamefont {Ryu}, \citenamefont {Furusaki},\ and\
  \citenamefont {Ludwig}}]{Schnyder_classification_free_2008}%
  \BibitemOpen
  \bibfield  {author} {\bibinfo {author} {\bibfnamefont {A.~P.}\ \bibnamefont
  {Schnyder}}, \bibinfo {author} {\bibfnamefont {S.}~\bibnamefont {Ryu}},
  \bibinfo {author} {\bibfnamefont {A.}~\bibnamefont {Furusaki}}, \ and\
  \bibinfo {author} {\bibfnamefont {A.~W.~W.}\ \bibnamefont {Ludwig}},\ }\href
  {\doibase 10.1103/PhysRevB.78.195125} {\bibfield  {journal} {\bibinfo
  {journal} {Phys. Rev. B}\ }\textbf {\bibinfo {volume} {78}},\ \bibinfo
  {pages} {195125} (\bibinfo {year} {2008})}\BibitemShut {NoStop}%
\bibitem [{\citenamefont {Kitaev}(2009)}]{Kitaev_classification_free_2009}%
  \BibitemOpen
  \bibfield  {author} {\bibinfo {author} {\bibfnamefont {A.}~\bibnamefont
  {Kitaev}},\ }\href {\doibase 10.1063/1.3149495} {\bibfield  {journal}
  {\bibinfo  {journal} {AIP Conf. Proc.}\ }\textbf {\bibinfo {volume} {1134}},\
  \bibinfo {pages} {22} (\bibinfo {year} {2009})}\BibitemShut {NoStop}%
\bibitem [{\citenamefont {Ryu}\ \emph {et~al.}(2010)\citenamefont {Ryu},
  \citenamefont {Schnyder}, \citenamefont {Furusaki},\ and\ \citenamefont
  {Ludwig}}]{Ryu_classification_free_2010}%
  \BibitemOpen
  \bibfield  {author} {\bibinfo {author} {\bibfnamefont {S.}~\bibnamefont
  {Ryu}}, \bibinfo {author} {\bibfnamefont {A.~P.}\ \bibnamefont {Schnyder}},
  \bibinfo {author} {\bibfnamefont {A.}~\bibnamefont {Furusaki}}, \ and\
  \bibinfo {author} {\bibfnamefont {A.~W.~W.}\ \bibnamefont {Ludwig}},\ }\href
  {http://stacks.iop.org/1367-2630/12/i=6/a=065010} {\bibfield  {journal}
  {\bibinfo  {journal} {New J. Phys.}\ }\textbf {\bibinfo {volume} {12}},\
  \bibinfo {pages} {065010} (\bibinfo {year} {2010})}\BibitemShut {NoStop}%
\bibitem [{\citenamefont {Budich}\ \emph {et~al.}(2018)\citenamefont {Budich},
  \citenamefont {Carlstr{\"o}m}, \citenamefont {Kunst},\ and\ \citenamefont
  {Bergholtz}}]{Budich_SPER_arXiv18}%
  \BibitemOpen
  \bibfield  {author} {\bibinfo {author} {\bibfnamefont {J.~C.}\ \bibnamefont
  {Budich}}, \bibinfo {author} {\bibfnamefont {J.}~\bibnamefont
  {Carlstr{\"o}m}}, \bibinfo {author} {\bibfnamefont {F.~K.}\ \bibnamefont
  {Kunst}}, \ and\ \bibinfo {author} {\bibfnamefont {E.~J.}\ \bibnamefont
  {Bergholtz}},\ }\href@noop {} {\bibfield  {journal} {\bibinfo  {journal}
  {arXiv preprint arXiv:1810.00914}\ } (\bibinfo {year} {2018})}\BibitemShut
  {NoStop}%
\end{thebibliography}


%%%%%%%%%%%%%%%%%%%%%%%%%%%%%%

\clearpage

\renewcommand{\thesection}{S\arabic{section}}
\renewcommand{\theequation}{S\arabic{equation}}
\setcounter{equation}{0}
\renewcommand{\thefigure}{S\arabic{figure}}
\setcounter{figure}{0}
\renewcommand{\thetable}{S\arabic{table}}
\setcounter{table}{0}
\makeatletter
\c@secnumdepth = 2
\makeatother

\onecolumngrid
\begin{center}
 {\large \textmd{Supplemental Materials:} \\[0.3em]
 {\bfseries 
 Exceptional rings in two-dimensional correlated systems with chiral symmetry
 }
 }
\end{center}

\setcounter{page}{1}

%%%%%%%%%%%%%%%%%%
\section{Symmetry and Green's function}
\label{sec: chi_symm_GF}
%%%%%%%%%%%%%%%%%%
We discuss a symmetry constraint on the effective Hamiltonian $H_{eff}$ defined by the single-particle Green's function:
%%%%%%%%%%%%%%%%%%
\begin{eqnarray}
G^{-1}(\omega+i\delta,\bm{k}) &=& (\omega+i\delta) \1  -h(\bm{k}) -\Sigma(\omega+i\delta,\bm{k}), \nonumber \\
                              &:=& (\omega+i\delta) \1 - H_{eff}(\omega,\bm{k}),
\end{eqnarray}
%%%%%%%%%%%%%%%%%%
where $h(\bm{k})$ denotes the Hermitian Hamiltonian matrix for free fermions. $\Sigma(\omega+i\delta,\bm{k})$ denotes the self-energy of frequency $\omega$ and momentum $\bm{k}$.

$G(\omega+i\delta,\bm{k})$ is a matrix whose Fourier transformed version is written as
%%%%%%%%%%%%%%%%%%
\begin{subequations}
\begin{eqnarray}
iG^R_{ab}(t) &=& \Theta(t>0) [\langle \hat{c}_a(t) \hat{c}^\dagger_b \rangle  + \langle \hat{c}^\dagger_b \hat{c}_a(t) \rangle  ], 
\end{eqnarray}
In a similar way, one can define $G(\omega-i\delta,\bm{k})$ whose Fourier transformed version is written as
%
%%%%%%%%%%%%%%%%%%
\begin{eqnarray}
-iG^A_{ab}(t) &=& \Theta(t<0) [\langle \hat{c}_a(t) \hat{c}^\dagger_b \rangle  + \langle \hat{c}^\dagger_b \hat{c}_a(t) \rangle  ]. 
\end{eqnarray}
%%%%%%%%%%%%%%%%%%
\end{subequations}
%%%%%%%%%%%%%%%%%%
Here for simplicity, we have introduced $a$ denoting the set of indices, $i$ and $n$; $\hat{c}_a:=\hat{c}_{in}$.
$\Theta(t>0)=1-\Theta(t<0)$ is the step function taking 1, 1/2, and 0 for $t>0$, $t=0$, and $t<0$, respectively.
In the following, we see that chiral symmetry for correlated systems results in Eq.~(\ref{eq: chiral_Heff_main})

%%%%%%%%%%%%%%%%%%
\subsection{Derivation of Eq.~(\ref{eq: chiral_Heff_main})}
%%%%%%%%%%%%%%%%%%
We show that the following constraint is imposed on $H_{eff}(\omega,\bm{k})$,
%%%%%%%%%%%%%%%%%%
\begin{eqnarray}
\label{eq: Heff chiral}
 H_{eff}(\omega,\bm{k})  &=& - U^\dagger_\Gamma H^\dagger_{eff}(-\omega,\bm{k}) U_\Gamma.
\end{eqnarray}
%%%%%%%%%%%%%%%%%%
when the many-body Hamiltonian is chiral symmetric:
%%%%%%%%%%%%%%%%%%
\begin{eqnarray}
 \hat{U}^\dagger_\Gamma \hat{H}^* \hat{U}_\Gamma &=& \hat{H}, \\
\hat{U}^\dagger_\Gamma \hat{c}_{jn} \hat{U}_\Gamma &=& \hat{c}^\dagger_{jm}U^\dagger_{\Gamma,mn}.
\end{eqnarray}
%%%%%%%%%%%%%%%%%%
Eq.~(\ref{eq: Heff chiral}) is obtained by combining Eqs.~(\ref{eq: GF chiral})~and~(\ref{eq: GF_Hermi}) which we see below.

We note that the chiral symmetry imposes the following constraints:
%%%%%%%%%%%%%%%%%%
\begin{eqnarray}
\label{eq: GF chiral}
G(\omega+i\delta)  &=& -U^\dagger_\Gamma G(-\omega-i\delta) U_\Gamma,
\end{eqnarray}
%%%%%%%%%%%%%%%%%%
The Fourier transformed version is written as 
%%%%%%%%%%%%%%%%%%
\begin{eqnarray}
G^R(t)  &=& -U^\dagger_\Gamma G^A(-t) U_\Gamma.
\end{eqnarray}
%%%%%%%%%%%%%%%%%%
This can be seen as follows. Because of chiral symmetry we have
%%%%%%%%%%%%%%%%%%
\begin{eqnarray}
\langle \hat{c}_a(t) \hat{c}^\dagger_b \rangle 
&=& \mathrm{Tr}[e^{-\beta \hat{H} } e^{it \hat{H} } \hat{c}_a e^{-it \hat{H} } \hat{c}^\dagger_b ] \nonumber \\
&=& \mathrm{Tr}[e^{-\beta \hat{U}^\dagger_\Gamma \hat{H}^* \hat{U}_\Gamma } e^{it \hat{U}^\dagger_\Gamma \hat{H}^* \hat{U}_\Gamma } \hat{c}_a e^{-it\hat{U}^\dagger_\Gamma H^* \hat{U}_\Gamma} \hat{c}^\dagger_b ] \nonumber \\
&=& \mathrm{Tr}[e^{-\beta  \hat{H}^* } e^{it \hat{H}^* } \hat{U}_\Gamma \hat{c}_a \hat{U}^\dagger_\Gamma e^{-it H^* } \hat{U}_\Gamma \hat{c}^\dagger_b  \hat{U}^\dagger_\Gamma ] \nonumber \\
&=& \mathrm{Tr}[e^{-\beta  \hat{H}^* } e^{it \hat{H}^* } U^\dagger_{\Gamma,aa'} \hat{c}^\dagger_{a'}  e^{-it H^* } \hat{c}_{b'}  U_{\Gamma,b'b} ] \nonumber \\
&=& U_{\Gamma,b'b} U^\dagger_{\Gamma,aa'} \mathrm{Tr}[ \hat{c}^\dagger_{b'} e^{-it H} \hat{c}_{a'} e^{it \hat{H} } e^{-\beta  \hat{H} }] \nonumber \\
&=& U_{\Gamma,b'b} U^\dagger_{\Gamma,aa'} \langle \hat{c}^\dagger_{b'} \hat{c}_{a'}(-t) \rangle.
\end{eqnarray}
%%%%%%%%%%%%%%%%%%
Here, we have used the following relations
%%%%%%%%%%%%%%%%%%
\begin{eqnarray}
 it \hat{H} &=& it \hat{U}_\Gamma \hat{H}^*\hat{U}^\dagger_\Gamma = \hat{U}_\Gamma it  H^* \hat{U}^\dagger_\Gamma, \\
\hat{U}_\Gamma \hat{c}_a \hat{U}^\dagger_\Gamma &=&  U^\dagger_{\Gamma,ab} \hat{c}^\dagger_{b},\\
 \langle N^*| A |M^*\rangle &=&  \langle M  |A^T |N\rangle,
\end{eqnarray}
%%%%%%%%%%%%%%%%%%
where $|N\rangle$ and $|M\rangle$ denote eigenstates of the many-body Hamiltonian.
In a similar way, we have the relations:
%%%%%%%%%%%%%%%%%%
\begin{eqnarray}
\langle \hat{c}^\dagger_b \hat{c}_a(t) \rangle &=& U^\dagger_{\Gamma,aa'} U_{\Gamma,b'b} \langle \hat{c}_{a'}(-t)  \hat{c}^\dagger_{b'} \rangle. 
\end{eqnarray}
%%%%%%%%%%%%%%%%%%
Thus, we have 
%%%%%%%%%%%%%%%%%%
\begin{eqnarray}
G^R(t) &=& -U^\dagger_\Gamma G^A(-t) U_\Gamma, 
\end{eqnarray}
%%%%%%%%%%%%%%%%%%
which results in Eq.~(\ref{eq: GF chiral}).

Now, we see another relation between $G^A$ and $G^R$.
For the Hermitian Hamiltonian, the following relations hold:
%%%%%%%%%%%%%%%%%%
\begin{eqnarray}
\label{eq: GF_Hermi_gen}
 G^*_{ba}(z) &=& G_{ab}(z^*),
\end{eqnarray}
%%%%%%%%%%%%%%%%%%
with $z\in \mathbb{C}$.
Here, $G_{ab}(z)$ is defined as 
%%%%%%%%%%%%%%%%%%
\begin{eqnarray}
G_{ab}(z) &:=& \sum_{NM} e^{\beta(\Omega -E_N)} \frac{e^{\beta(E_N-E_M)}+1}{z+E_N-E_M} \langle N | \hat{c}_a |M \rangle \langle M | \hat{c}^\dagger_b | N\rangle, \\
\end{eqnarray}
%%%%%%%%%%%%%%%%%%
with $e^{\beta(\Omega)}:=\sum_N e^{-\beta E_N }$ and the eigenenergy $E_N$ for the many-body Hamiltonian ($E_N\in \mathbb{R}$). 
In particular, for $z=\omega+i\delta$, Eq.~(\ref{eq: GF_Hermi}) indicates that 
%%%%%%%%%%%%%%%%%%
\begin{eqnarray}
\label{eq: GF_Hermi}
 G^\dagger_{ab}(\omega+i\delta) &=& G_{ab}(\omega-i\delta).
\end{eqnarray}
%%%%%%%%%%%%%%%%%%
Eq.~(\ref{eq: GF_Hermi_gen}) can be seen with the following straightforward calculation:
%%%%%%%%%%%%%%%%%%
\begin{eqnarray}
G^*_{ba}(z) 
&&=\sum_{NM} e^{\beta(\Omega -E_N)} \frac{e^{\beta(E_N-E_M)}+1}{(z+E_N-E_M)^*} \langle N | \hat{c}_b |M \rangle^* \langle M | \hat{c}^\dagger_a |N\rangle^* \nonumber \\
&&=\sum_{NM} e^{\beta(\Omega -E_N)} \frac{e^{\beta(E_N-E_M)}+1}{z^*+E_N-E_M}   \langle N | \hat{c}_a |M \rangle   \langle M | \hat{c}^\dagger_b |N\rangle   \nonumber \\
&&=G_{ab}(z^*).
\end{eqnarray}
%%%%%%%%%%%%%%%%%%

Combining Eqs.~(\ref{eq: GF chiral})~and~(\ref{eq: GF_Hermi_gen}) yields Eq.~(\ref{eq: GF chiral}).
%%%%%%%%%%%%%%%%%%
Therefore, we obtain the relation~(\ref{eq: Heff chiral}).

%%%%%%%%%%%%%%%%%%
\section{Enhancement of the specific heat due to Fermi planes}
\label{sec: cv}
%%%%%%%%%%%%%%%%%%

We here show that the Fermi planes accompanying the SPERs enhance the specific heat which is computed by the relation $C=d\langle \hat{H}\rangle/d T$.
%%%%%%%%%%%%%%%%%%%%%%%%%%
\begin{figure}[!h]
\begin{minipage}{0.475\hsize}
\begin{center}
\includegraphics[width=\hsize,clip]{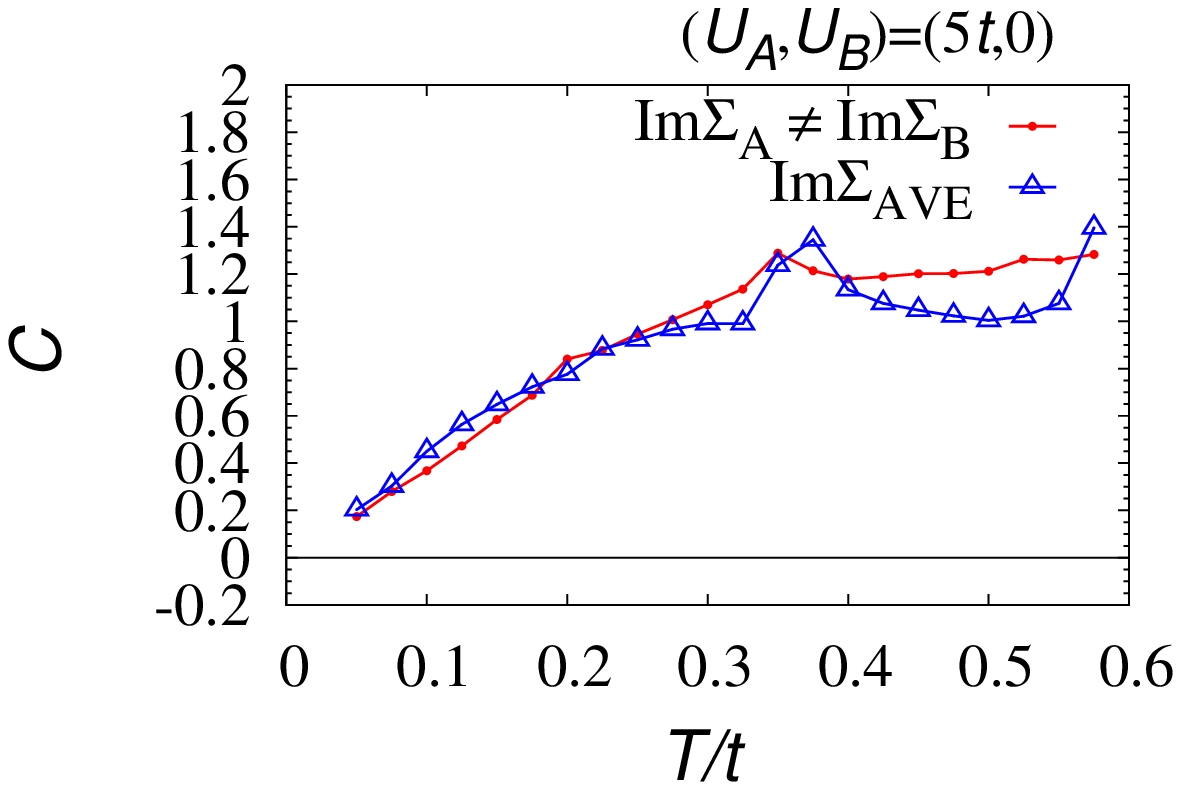}
\end{center}
\end{minipage}
\begin{minipage}{0.475\hsize}
\begin{center}
\includegraphics[width=\hsize,clip]{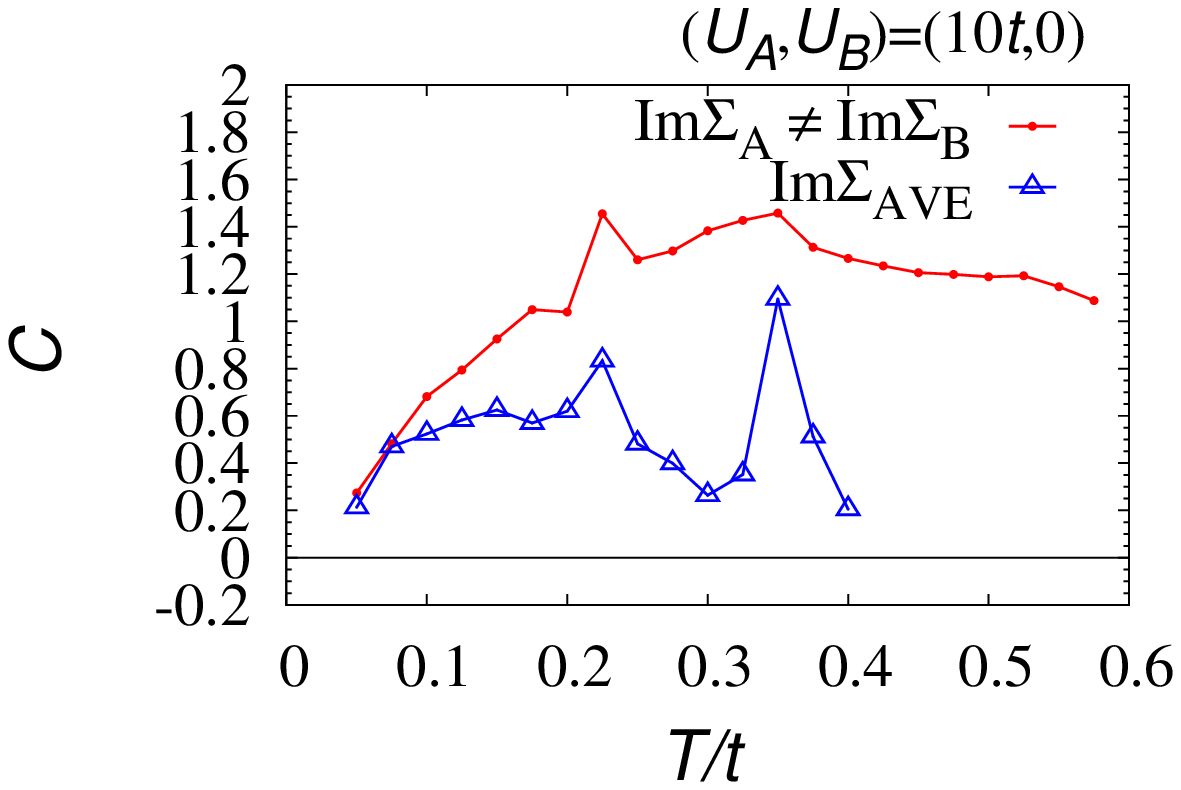}
\end{center}
\end{minipage}
\caption{(Color Online).
Specific heat as functions of temperature.
The left (right) panel represents data for $(U_A,U_B)=(5t,0)$ and $[(10t,0)]$, respectively.
The blue line is obtained by setting $\mathrm{Im} \Sigma_A=\mathrm{Im} \Sigma_B=\mathrm{Im} (\Sigma_A+\Sigma_B)/2$ so that $d_3$ becomes zero.
In the case of $(U_A,U_B)=(10t,0)$, we could not compute the specific heat accurately for $T>0.4t$.
}
\label{fig: Cv}
\end{figure}
%%%%%%%%%%%%%%%%%%%%%%%%%
In Fig.~\ref{fig: Cv}, the specific heat is plotted as a functions of the temperature, which is shown as a red line. 
For a comparison, we also plot the specific heat by supposing that both for $A$ and $B$ sublattices, the imaginary part of the self-energy takes the average value $\mathrm{Im}[\Sigma_A(\omega+i\delta)+\Sigma_B(\omega+i\delta)]/2$, which is shown as a blue line. In this figure, one can see that the specific heat (read line) is enhanced, compare to the one shown as a blue line both for $U_A=5t$ and $U_A=10t$.

%%%%%%%%%%%%%%%%%%
\section{
Energy dispersion for the case beyond non-Hermitian Dirac Hamiltonians
}
\label{sec: BeyondDH_E}
%%%%%%%%%%%%%%%%%%
We here discuss the energy dispersion of the following non-Hermitian Hamiltonian
%%%%%%%%%%%%%%%%%%
\begin{eqnarray}
H_{eff}(0,\bm{k}) &=& b_1(\bm{k}) \tau_1 \sigma_3 +b_2(\bm{k}) \tau_2 \sigma_3 + V\tau_3 \sigma_1 +id_3\tau_3\sigma_3, \nonumber 
\end{eqnarray}
%%%%%%%%%%%%%%%%%%
which is discussed in the main text. $b_1$ and $b_2$ are defined in Eq.~(\ref{eq: bs_honeycomb}). Here $V$ and $d_3$ are real constants.
%%%%%%%%%%%%%%%%%%%%%%%%%
\begin{figure}[!h]
\begin{minipage}{0.45\hsize}
\begin{center}
\includegraphics[width=\hsize,clip]{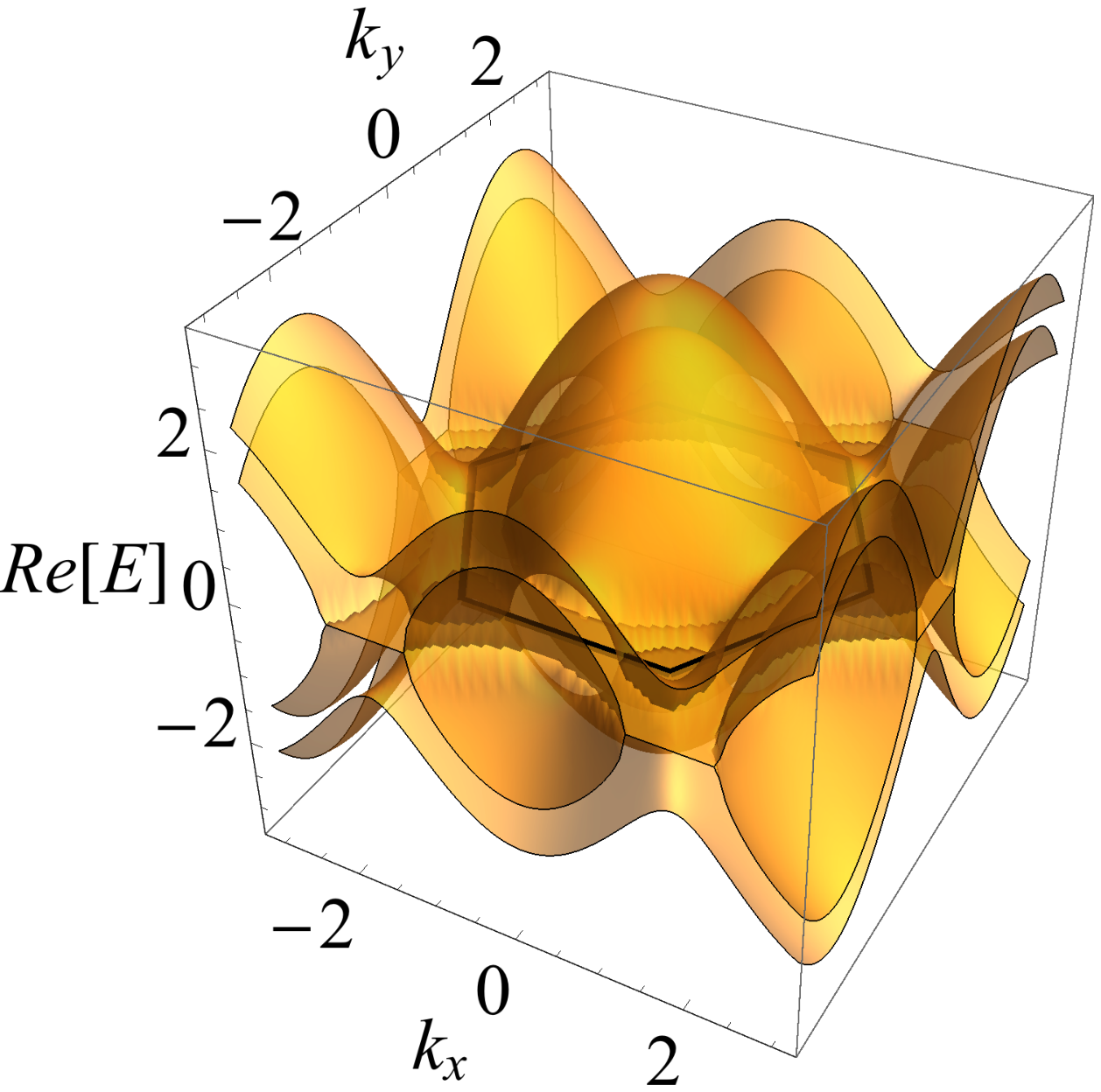}
\end{center}
\end{minipage}
\begin{minipage}{0.45\hsize}
\begin{center}
\includegraphics[width=\hsize,clip]{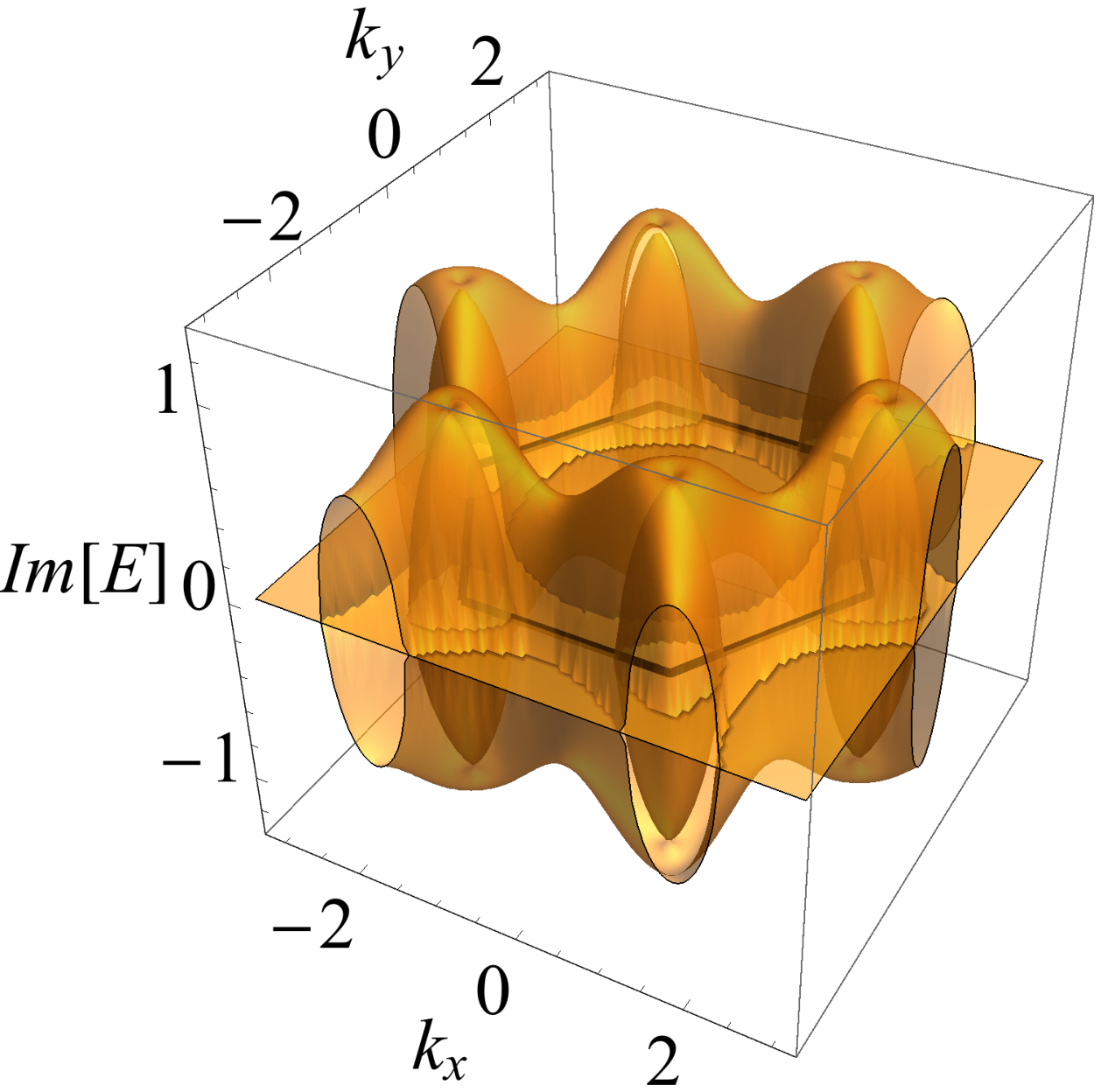}
\end{center}
\end{minipage}
\caption{(Color Online).
Energy spectrum for 
$
\left(
\begin{array}{ccc}
t &V& d
\end{array}
\right)
=
\left(
\begin{array}{ccc}
1 &0.3& 1
\end{array}
\right)
$. 
The left (right) panel represent the real (imaginary) part.
Black hexagons in these figures illustrate the BZ.
}
\label{fig: BeyondDH_E}
\end{figure}
%%%%%%%%%%%%%%%%%%%%%%%%%
Energy dispersion for $V=0.3$ is shown in Fig.~\ref{fig: BeyondDH_E}. Because of the finite value of $V$, the Hamiltonian cannot be expanded by Dirac matrices satisfying anti-commutation relations.
Correspondingly, each energy band splits. However, intriguingly, we still can observe the Fermi planes; in the left panel of Fig.~\ref{fig: BeyondDH_E}, one can find the region where the real-part of the energy vanishes, which is due to the robustness of SPERs.

\end{document}